\documentclass[11pt]{article}
\usepackage{bbm}
\usepackage[final]{graphics}
\usepackage{amsmath}
\usepackage{amsfonts,amsbsy}
\usepackage{amssymb}
\usepackage{bm}
\usepackage{color}

\def\empile#1\over#2{\mathrel{\mathop{\kern 0pt#1}\limits_{#2}}}
\def\bs{\boldsymbol}

\def\TODO#1{}

\def\k{{\boldsymbol k}}

\renewcommand{\d}{\ensuremath{\mathrm{d}}}

\newcommand{\x}{\bm{x_{\perp}}}
\newcommand{\p}{\bm{p_{\perp}}}
\newcommand{\ka}{\bm{k_{\perp}}}
\newcommand{\y}{\bm{y_{\perp}}}
\newcommand{\intp}{\int\frac{\d^2\p}{(2\pi)^2}}
\newcommand{\ma}[1]{{\mathcal{#1}}}

\begin{document}

%\title{\bf The spectrum of fluctuations over\\ 
%the classical Yang-Mills background\\
%in high energy heavy ion collisions}

\title{\bf Fluctuations of the initial color fields\\
in high energy heavy ion collisions}
\author{Thomas Epelbaum, Fran\c cois Gelis}
\maketitle
 \begin{center}
  Institut de Physique Th\'eorique (URA 2306 du CNRS)\\
   CEA/DSM/Saclay, 91191 Gif-sur-Yvette Cedex, France
 \end{center}

\begin{abstract}
  In the Color Glass Condensate approach to the description of high
  energy heavy ion collisions, one needs to superimpose small random
  Gaussian distributed fluctuations to the classical background field,
  in order to resum the leading secular terms that result from the
  Weibel instability, that would otherwise lead to pathological
  results beyond leading order. In practical numerical simulations,
  one needs to know this spectrum of fluctuations at a proper time
  $\tau \ll Q_s^{-1}$ shortly after the collision, in the
  Fock-Schwinger gauge $\ma{A}^\tau=0$.

  In this paper, we derive these fluctuations from first principles,
  by solving the Yang-Mills equations linearized around the classical
  background, with plane wave initial conditions in the remote past.
  We perform the intermediate steps in light-cone gauge, and we
  convert the results to the Fock-Schwinger gauge at the end.  We
  obtain simple and explicit formulas for the fluctuation modes.
\end{abstract}

\section{Introduction}
One of the outstanding theoretical problems in high energy heavy ion
collisions is the understanding from first principles of the pressure
isotropization and possibly the thermalization of the gluonic matter
produced in these collisions.

From RHIC and LHC data, there is ample evidence that the expansion and
cooling of this matter is well described by relativistic hydrodynamics
\cite{Teane1,Teane2,Ollit2,OllitG1} with a very small viscosity
(characterized by a viscosity to entropy density ratio, $\eta/s$, that
is fairly close to the value $1/4\pi$ obtained in the strong coupling
limit of some QCD-like theories \cite{PolicSS1}, and that has been
conjectured to be a lower bound). This good agreement also suggests
that the anisotropy between the transverse and longitudinal (with
respect to the collision axis) pressures is not too large, because
otherwise the viscous corrections could be important and spoil this
agreement. However, understanding from first principles why the
hydrodynamics models work so well has proven very challenging until
now.

Moreover, there is also a vast amount of data, ranging from Deep
Inelastic Scattering to proton-nucleus and nucleus-nucleus collisions,
supporting the idea of gluon saturation in high energy collisions
involving hadrons or nuclei \cite{BlaizG1,Lappi6,GelisIJV1}. In this
regime, the gluon density in the projectiles becomes very large,
leading to important non-linear corrections in the evolution of the
gluon distribution with energy.  These nonlinear effects dynamically
generate a dimensionful scale, the saturation momentum $Q_s$, that
controls the scattering \cite{GriboLR1,MuellQ1}. The gluon occupation
number is non-perturbatively large, of order $1/\alpha_s$, for
transverse momenta below $Q_s$, and decreases rapidly above this
scale. The saturation momentum increases with the energy of the
collision, to reach values of order $Q_s\approx 1-2~$GeV for nuclei at
LHC energies. Since the value of the strong coupling $\alpha_s$ at
such scales is around $\alpha_s\approx 0.3$, one may expect to be able
to describe these collisions in the Color Glass Condensate effective
theory \cite{IancuLM3,IancuV1,GelisIJV1,Gelis15}, that describes the
physics of gluon saturation at weak coupling\footnote{Note that
  ``weakly coupled'' does not imply ``weakly interacting'', nor
  ``perturbative'', because of the non-perturbatively large gluon
  occupation number.}.

For this reason, the CGC appears to be a well suited framework in
order to try to explain the early isotropization of the system. The
state of the system just after such a collision has been calculated at
Leading Order in $\alpha_s$ in the CGC framework \cite{KovneMW2}, and
one finds that its energy-momentum tensor is very anisotropic, with a
negative longitudinal pressure exactly opposite to the energy density
(a trivial consequence of the fact that the chromo-electric and
chromo-magnetic fields are parallel to the collision axis just after
the collision \cite{LappiM1}). At leading order, the subsequent time
evolution never leads to the isotropization of the stress tensor
\cite{KrasnV1,KrasnV3,KrasnNV2,Lappi1,FukusG1}. But it has been
noticed long ago that the CGC result at leading order is insufficient,
because of the existence of instabilities in the classical solutions
of the Yang-Mills equations \cite{RomatV1,RomatV2,RomatV3,KunihMOST1}:
some of the higher order (in $\alpha_s$) corrections grow
exponentially fast with time, and soon become larger than the leading
order they are supposed to correct. These instabilities are the
manifestation in the CGC framework of the well known Weibel
instabilities in plasmas with an anisotropic particle distribution
\cite{Mrowc2,Mrowc3,RomatS1,RomatS2}. Moreover, a lot of work suggests
that these instabilities could play an important role in driving the
system towards isotropization and local thermal equilibrium
\cite{RebhaS1,RebhaSA1,ArnolLM1,ArnolLMY1,ArnolM3,KurkeM1,KurkeM2,BodekR1,AttemRS1}.

In the CGC formalism, these instabilities spoil the naive estimates of
the order of magnitude of contributions
\cite{GelisV2,GelisLV3,GelisLV2}, since these estimates usually keep
track only of the powers of $\alpha_s$, implicitly assuming that all
the numerical prefactors remain of order unity at all times. In the
presence of unstable modes, this is no longer true: some of these
coefficients will grow exponentially in time, leading to secular
divergences when the time goes to infinity -- and making the ordinary
loop expansion useless after a finite time of order $Q_s^{-1}$.  An
improved power counting that tracks these fast growing terms was
proposed in ref.~\cite{DusliGV1}, and it was shown
\cite{GelisLV2,DusliGV1,FukusGM1} that one can resum the fastest
growing terms by superimposing random Gaussian fluctuations to the
initial condition of the classical Yang-Mills equations, and then
averaging over these fluctuations. Thanks to this resummation, one
completely tames the secular terms, and the validity of the resummed
result is extended to larger times.

As a proof of concept, this resummation was implemented numerically in
the case of a $\phi^4$ scalar field theory. Although very different
from a Yang-Mills theory in many respects, this theory has several
similar features: it is scale invariant in 3+1 dimensions at the
classical level, and its classical solutions have instabilities (here
due to parametric resonance). It was shown in
Refs.~\cite{DusliEGV1,EpelbG1,DusliEGV2} that after performing the
Gaussian average over the fluctuations of the initial classical field,
the system evolves towards the equilibrium equation of state, and that
its transverse and longitudinal pressures become equal in the case of
a system expanding in the longitudinal direction.

Moreover, the origin of this resummation scheme (and in particular the
fact that it includes the exact NLO result) completely prescribes the
ensemble of these fluctuations: their spectrum can be obtained by
computing a 2-point correlator in the presence of a non-trivial
background field (the solution of Yang-Mills equations at leading
order). From the analysis of next-to-leading order corrections done in
\cite{GelisLV3}, this 2-point function can be constructed as follows
\begin{eqnarray}
{\cal G}^{\mu a,\nu b}(x,y)
&=&
\sum_{\lambda,c}\int \frac{\d^3\k}{(2\pi)^3 2k}\;
a^{\mu a}_{\k \lambda c}(x)\,a^{\nu b *}_{\k \lambda c}(y)\; ,
\label{eq:fluct000}
\end{eqnarray}
where $a^{\mu a}_{\k \lambda c}(x)$ is the solution of the Yang-Mills
equations linearized around the classical CGC background, whose
initial condition at $x^0=-\infty$ is a plane wave of momentum $\k$,
polarization $\lambda$ and color $c$. However, this
calculation has never been done so far.

In ref.~\cite{DusliGV1}, an alternative way of computing these
fluctuations was proposed, based on the existence of an inner product
between pairs of these fluctuations (written here in terms of the
proper time $\tau$, the rapidity $\eta$ and the transverse coordinate
$\x$),
\begin{eqnarray}
\big(a_1\big|a_2\big)&\equiv&
-i\int \d^2 \x \d \eta \;g_{\mu\nu}\delta_{ab}
\Big(
a_1^{\mu{{a}*}}(\tau,\x,\eta)
e_2^{\nu{b}}(\tau,\x,\eta)
\nonumber\\
&&\qquad\qquad\qquad\qquad-
e_1^{\mu{a}*}(\tau,\x,\eta)
a_2^{\nu{b}}(\tau,\x,\eta)\Big)\;.
\label{eq:centralps}
\end{eqnarray}
$e^\mu$ denotes the electrical field associated to the gauge
potential $a^\mu$, defined as~:
\begin{equation}
e^i\equiv \tau\partial_\tau a_i\quad,\quad 
e^\eta \equiv \tau^{-1}\partial_\tau a_\eta\; .
\end{equation}
The above inner product is conserved when the fluctuations evolve over
the classical background. It is also easy to check that the modes
obtained by evolving plane waves from the remote past form an
orthonormal (with respect to the above inner product) basis of the
vector space of fluctuations. It was then suggested that one may avoid
solving the linearized equations of motion for the fluctuations from
the remote past, and that it could be sufficient to find a complete
set of modes that obey the equations of motion locally near a proper
time $\tau>0$ just after the collision has taken place, provided that
this set of modes also form an orthonormal basis in the above
sense. Solving this alternate problem is simpler because one needs
only to find {\sl local} solutions of the linearized equations of
motion, instead of global solutions with prescribed initial conditions
at $x^0=-\infty$.

The reasoning in \cite{DusliGV1} was that if one knows a set of
orthonormal modes at the time $\tau$, even if it is not the same set
as the one originating from the plane waves, it would generate the
same Gaussian ensemble of fluctuations provided that the two basis can
be related by a unitary transformation. And it is also clear that
unitary transformations preserve the inner product defined in
eq.~(\ref{eq:centralps}). It turns out that there is a caveat in this
argument: there are also non-unitary transformations that preserve the
inner product. Such a transformation, when applied to a basis of
fluctuation modes, will leave all the inner products unchanged (and
thus transform an orthonormal basis into another orthonormal basis)
but it will lead to a different Gaussian ensemble of fluctuations.

A very simple example of such a transformation is to multiply all the
electrical fields by a constant $\lambda$, while at the same time
dividing the gauge potentials by the same constant\footnote{More
  generally, one may note that the inner product defined in
  eq.~(\ref{eq:centralps}) is the complex version of a simplectic
  product. It is invariant under all the canonical transformations of
  the fields and their conjugate momenta, that form a superset of the
  unitary transformations (where one would apply the same unitary
  rotation both to the gauge potentials and to the electrical
  fields).}. Obviously this transformation does not change the inner
product defined in eq.~(\ref{eq:centralps}), but it multiplies the
variance of the set of Gaussian fluctuations by $\lambda^2$ for the
electrical fields, and by $\lambda^{-2}$ for the gauge
potentials. Given the existence of these transformations, one cannot
be sure that the set of mode functions obtained in \cite{DusliGV1}
leads to the correct\footnote{In the special case where the background
  field vanishes, then the modes found in \cite{DusliGV1} are indeed
  the correct ones, as they can easily be related to plane waves. The
  issue exists only for the case of a non trivial (i.e. non pure
  gauge) background field.}  fluctuations.  Instead, they should be
constructed by evolving the plane waves from $x^0=-\infty$.

In the present paper, we reconsider this question by going back to the
original definition of the 2-point function that controls the Gaussian
spectrum of fluctuations, i.e. the eq.~(\ref{eq:fluct000}). Using a
gauge fixing inspired from ref.~\cite{BlaizM1}, we explicitly solve
the linearized Yang-Mills equations over the leading order classical
background field, with plane waves as the initial condition in the
remote past. We obtain rather simple analytical expressions for these
solutions, at a proper small positive time $\tau\ll Q_s^{-1}$
(i.e. just after the collision). We provide the results in the
Fock-Schwinger gauge that is commonly employed in the numerical
resolution of the Yang-Mills equations, for a choice of quantum
numbers which is appropriate for a numerical implementation on a
lattice with a fixed spacing in the rapidity $\eta$ (as opposed to a
discretization with a fixed spacing in the longitudinal coordinate
$z$).

The paper is organized as follows. In the section
\ref{sec:background}, we recall some well known results for the
solution of the classical Yang-Mills equations in the presence of the
two color currents that describe the colliding nuclei. Most of the
section is devoted to summarizing the derivation of this solution in
the $\ma{A}^-=0$ gauge, originally performed in ref.~\cite{BlaizM1},
and on the gauge transformation that one must perform in order to
eventually obtain the result in the Fock-Schwinger gauge. In the
section \ref{sec:fluct}, we follow a similar strategy in order to
solve the linearized Yang-Mills equations for a small perturbation
propagating over this background field. The calculation is subdivided
in several stages, corresponding to the successive encounters of the
fluctuation with the two nuclei, followed by a final gauge
transformation to go from the $\ma{A}^-=0$ gauge to the Fock-Schwinger
gauge. The impatient reader may find the final result in
eqs.~(\ref{eq:finalresult}). The section \ref{sec:concl} is devoted to
concluding remarks, and some more technical material is relegated into
several appendices.

\section{Classical background field}
\label{sec:background}
\subsection{General setup of the problem}
Before going into the details of our calculation, let us state the
problem we need to solve, by listing the equations of motion and
current conservation constraints that must be satisfied, as well as
the boundary conditions that are appropriate in applications to heavy
ion collisions. Here, we list these equations in a generic form which
is valid in any gauge. As we shall see later, specific gauge choices
may lead to some simplifications.

At leading order in the CGC framework, inclusive observables can be
expressed in terms of a gauge field that obeys the classical
Yang-Mills equations, and that vanishes in the remote past
(i.e. before the collision)~:
\begin{align}
\left[\ma{D}_\mu,\ma{F}^{\mu\nu}\right]=\null& J^{\nu}\;,&
\left[\ma{D}_\mu,J^{\mu}\right]=\null&0\;, \label{eq:YMLO}\\
\lim\limits_{t\to-\infty}\ma{F}^{\mu\nu}=\null&0\;,&
\lim\limits_{t\to-\infty}J^{\nu}=\null&
\delta^{\nu-}\rho_1+
\delta^{\nu+}\rho_2\;.
\end{align}
On the left are the equations obeyed by the gauge potential (or
equivalently the field strength ${\cal F}^{\mu\nu}$), and on the right
are the equations satisfied by the external current $J^\nu$. In the
remote past, it is given simply in terms of the two functions $\rho_1$
and $\rho_2$ that represent the color charge distribution in the two
nuclei before the collision. However, since its conservation equation
involves a covariant derivative, this current can be modified during
the collision by the radiated gauge fields. This means that in
general, one must view the eqs.~(\ref{eq:YMLO}) as coupled equations.
This problem has been solved long ago in \cite{KovneMW2,KovneMW1}. We
just briefly remind the reader of the solution in the rest of this
section, and we also discuss an alternate way of solving these
equations that has been proposed in \cite{BlaizM1}.

When extending the CGC to next-to-leading order, one needs to study
small perturbations to the gauge field, more specifically those that
behave as plane waves before the collision. Because the gauge field is
entangled with the current via the conservation equation, this in
general leads to a small perturbation to the current as
well\footnote{This has a simple physical interpretation~: in
  non-abelian gauge theories, the incoming plane wave $a^\mu$ carries
  a color. A quantum from this wave can be absorbed by one of the
  charges that contribute to the current $J^\mu$, thereby altering its
  color, and therefore changing the current itself.}. A linearization
of the above equations around the LO solution gives~:
\begin{align}
\left[\ma{D}_{\mu},\left[\ma{D}^{\mu},a^{\nu}\right]
-\left[\ma{D}^{\nu},a^{\mu}\right]\right]
-ig\left[\ma{F}^{\nu\mu},a_{\mu}\right]=\null&j^{\nu}\;,&
\lim\limits_{t\to-\infty}a^{\mu}=\null&\epsilon^{\mu} e^{ik\cdot x}\;,
\end{align}
and
\begin{align}
\left[\ma{D}_{\mu},j^{\mu}\right]-ig\left[a_\mu,J^{\mu}\right]=\null&0\;,&
\lim\limits_{t\to -\infty}j^{\nu}=\null&0\;.
\end{align}
Note that the change $j^\nu$ to the current must vanish in the remote
past, since this is before the current could possibly have been
altered by the plane wave. Depending on the gauge choice, the bracket
$\left[a_\mu,J^{\mu}\right]$ may vanish and therefore the perturbation
of the current is identically $0$. (But in the collision of two
projectiles, the current has both non-vanishing $J^+$ and $J^-$
components, and none of the light-cone gauges can eliminate this term
completely).

\subsection{Reminder of standard results}
In the CGC description of heavy ion collisions, the gauge fields are
driven by two color currents $J_1^-(x^+,\x)$ and $J_2^+(x^-,\x)$ that
describe the color carried by the fast partons of the two
projectiles. These currents are proportional to delta distributions
$\delta(x^+)$ and $\delta(x^-)$, respectively.  Because of the
presence of these singular sources in the classical Yang-Mills
equations, one starts the numerical resolution of the field equations
of motion slightly above the forward light-cone, at some small proper
time $\tau>0$. The evolution of the fields is thus free of these
singular sources, but the drawback is that one must know the initial
value of the gauge potentials and electrical fields at the starting
time $\tau$.

These initial conditions were first obtained in
refs.~\cite{KovneMW2,KovneMW1} from the known values of these fields
below the light-cone, by a matching procedure that amounts to
requesting that all the singularities cancel from the solution.  At a
proper time $\tau=0^+$ immediately after the collision, the initial
conditions read\footnote{The formulas written without explicit color
  indices, like eqs.~(\ref{eq:mvfields}-\ref{eq:A1}) in this section,
  are valid in any representation of the SU(N) algebra.}
\begin{eqnarray}
\ma{A}^{\tau}_{_{\rm FS}}(\x)&=&0\ \ \mbox{(gauge condition)}
\nonumber\\
\ma{A}^{i}_{_{\rm FS}}(\x)&=&\alpha_1^{i}(\x)+\alpha_2^{i}(\x)
\nonumber\\
\ma{A}^{\eta}_{_{\rm FS}}(\x)&=&\frac{ig}{2} \Big[\alpha_1^{i}(\x),\alpha_2^{i}(\x)\Big]\;,
\label{eq:mvfields}
\end{eqnarray}
where the fields $\alpha_{1,2}$ are the solutions in light-cone gauge
of the classical Yang-Mills equations for a single projectile. For the
projectile moving in the $-z$ direction, we have
\begin{eqnarray}
\alpha_1^i(x^+,\x)&=&
\frac{i}{g}\,\ma{U}_{1}^\dagger(x^+,\x)\partial^i\ma{U}_{1}(x^+,\x)\; ,
\nonumber\\
\ma{U}_{1}(x^+,\x)&=&
{\rm T} \,e^{ig\int_{-\infty}^{x^+}\d z^+\,A_1^-(z^+,\x)}\;,
\label{eq:U1}
\end{eqnarray}
where $A_1^{-}$ (which can be viewed as the gauge potential of that
nucleus in Lorenz gauge) is related to the corresponding color current
by
\begin{equation}
-{\bs\nabla}_\perp^2\; A_1^-(x^+,\x)= J^-_1(x^+,\x)\; .
\label{eq:A1}
\end{equation}
A similar set of equations relates the field $\alpha_2^i$ to the color
current $J_2^+$ of the second nucleus. Note that the $x^\pm$
dependence of $\alpha_{1,2}^i$ is relevant only inside the support of
the color currents. Outside of these (infinitesimal) regions along the
light-cones, the Wilson lines $\ma{U}_{1,2}$ depend only on the
transverse coordinate $\x$. This is why in eq.~(\ref{eq:mvfields}),
all the fields have only a transverse dependence. It is sometimes 
useful in intermediate steps of the calculations to extend the support
of these currents to a small but finite range $0<x^\pm<\epsilon$. The
limit $\epsilon\to 0^+$ is always taken at the end of the
calculations, and the final answers will all be given for color
currents that have an infinitesimal support.

In ref.~\cite{BlaizM1}, the initial conditions (\ref{eq:mvfields})
have been rederived by doing all the intermediate calculations in
light-cone gauge, where crossing the light-cones that support the
color currents can be handled more easily. Since this is also the
gauge choice that we will adopt for the intermediate steps of our
calculation of the fluctuations, we reproduce the main steps of
\cite{BlaizM1} in the rest of this section, in order to outline its
key features.

\subsection{Solution in the global light-cone gauge $\ma{A}^-=0$}
In the first derivation of the initial fields of
eqs.~(\ref{eq:mvfields}), different light-cone gauges were used for
describing the two projectiles before they collide, by exploiting the
fact that there is no causal contact between them until the collision.

The main modification introduced in \cite{BlaizM1} is to use a unique
light-cone gauge, that is employed globally to treat the two
projectiles. In this work, we will choose the $\ma{A}^-=0$ gauge
condition for this purpose. This choice breaks the symmetry between
the two nuclei. For the nucleus moving in the $+z$ direction, the
solution of the Yang-Mills equations in Lorenz gauge,
\begin{equation}
-{\bs\nabla}_\perp^2\; A_2^+(x^-,\x)= J^+_2(x^-,\x)\; ,
\label{eq:A2}
\end{equation}
fulfills the light-cone gauge condition $\ma{A}^-=0$ and
therefore does not need to be transformed further. Therefore, we just
take
\begin{equation}
\ma{A}_2^+ = A_2^+\; .
\end{equation}

\begin{figure}[htbp]
\begin{center}
\resizebox*{!}{3.5cm}{\includegraphics{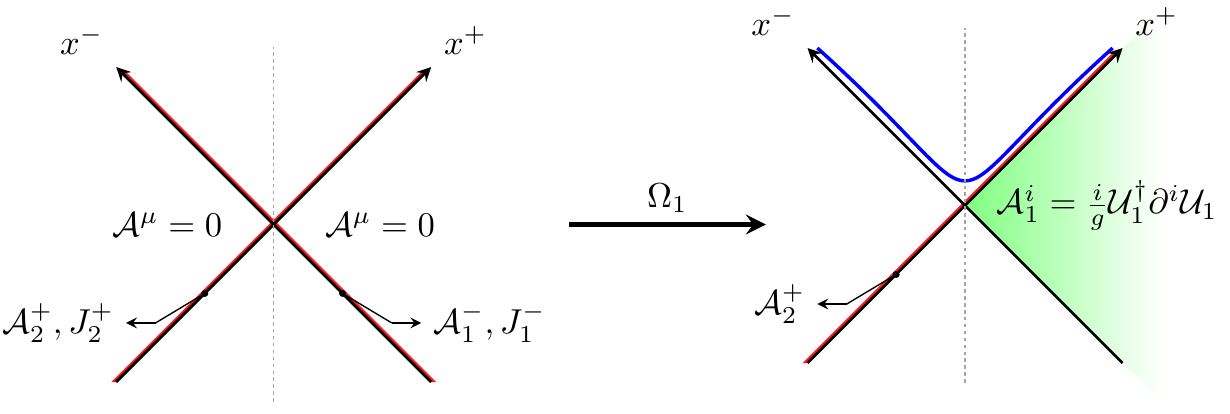}}
\end{center}
\caption{\label{fig:gauget}The gauge transformation that transforms
  the Lorenz gauge field ${A}^{-}_1$ into the light-cone gauge field
  $\ma{A}^{i}_1$. The second nucleus is unaffected by this
  transformation.}
\end{figure}

This is not the case for the nucleus 1, whose gauge potential in
Lorenz gauge has a non-zero minus component. We thus need to perform a
gauge transformation,
\begin{equation}
\ma{A}^{\mu}_1 
=
\Omega_1^\dagger {A}^{\mu}_1 \Omega_1
+
\frac{i}{g} \Omega_1^{\dagger}\partial^{\mu}\Omega_1\; .
\end{equation}
The gauge transformation $\Omega_1$ has to be chosen so that it
eliminates the minus component, and it turns out that it should be
equal to the Wilson line $\ma{U}_1$ introduced earlier in
eq.~(\ref{eq:U1}). After the transformation, the non-zero components
of the field $\ma{A}_1^\mu$ are the transverse ones, and moreover they
have the form of a transverse pure gauge\footnote{Note that this is
  not a global pure gauge, since the gauge rotation $\Omega_1$ has
  different values at $x^+<0$ and $x^+>\epsilon$.}
\begin{equation}
\ma{A}^{i}_1 
=
\frac{i}{g} \ma{U}_1^\dagger\partial^{i}\ma{U}_1\; .
\end{equation}
(This field is of course identical to the $\alpha_1^i$ defined in
eq.~(\ref{eq:U1}). We will denote it ${\cal A}_1^i$ here for
consistency with the notation used for the field of the nucleus 2, and
to stress the fact that we are now in a different gauge). Note also
that the gauge transformation $\Omega_1$ has no incidence on the field
of the first nucleus, since it differs from the identity only at
$x^+>0$.  The figure \ref{fig:gauget} summarizes the structure of
the gauge potentials before and after the gauge transformation $\Omega_1$.

A legitimate question that arises is what is the advantage in treating
the two nuclei in such an asymmetric fashion? The reason is only
technical: many calculations turn out to be simpler in this mixed
description. In order to determine the fields just after the collision
(i.e. on the blue surface on the right side of figure
(\ref{fig:gauget})), one can independently study what happens on its
left and right branches. Indeed, causality prevents the field that
travels on the left side of the light-cone from interacting with the
field that travels on the right side\footnote{This is not true anymore
  in the forward light-cone, i.e. after the collision, where the
  fields on the left and on the right can now interact. Therefore,
  this simplification can only be used to calculate the fields on the
  surface $\tau=0^+$.} (they travel through regions that are separated
by space-like intervals).  The result is the same on the two branches
and for infinitesimal $x^+$ or $x^-$, it is given by\footnote{ Here,
  we have written all the color indices explicitly to avoid possible
  ambiguities. For instance, the second equation could equivalently be
  written as
  \begin{equation*}
    \ma{A}^{i}(\x)\equiv \ma{A}^{ia}(\x) t^a = \ma{U}_2 \ma{A}_1^{i}(\x) \ma{U}_2^\dagger\; ,
  \end{equation*}
  where all the objects in the right hand side should be in the same
  representation as the generators $t^a$.}~\cite{BlaizM1}
\begin{eqnarray}
\partial^-\ma{A}^{+a}(\x)&=&\left(\partial^i\ma{U}_2(\x)\right)_{ab}\ma{A}_1^{ib}(\x)
\nonumber\\
\ma{A}^{ia}(\x)&=&\ma{U}_{2ab}(\x)\ma{A}_1^{ib}(\x)
\nonumber\\
\ma{A}^{\pm a}(\x)&=&0\;.
\label{eq:fieldsLC}
\end{eqnarray}
One sees that $\ma{A}^\mu$ only depends on $\x$ on the blue surface of
figure (\ref{fig:gauget}), in the limit where this surface becomes
infinitesimally close to the forward light-cone. Note also that this
solution is not quite symmetric between the nuclei 1 and 2. Indeed,
$\partial^-\ma{A}^{+}$ is non zero, while $\partial^+\ma{A}^{-}$ is
identically zero  by virtue of the light-cone gauge condition.

\subsection{Transformation into the Fock-Schwinger gauge} 
\label{sec:Fsgaugebcg}
Above the forward light-cone, analytical solutions of the classical
Yang-Mills equations are not known, and one must resort to numerical
techniques. In principle, it would be perfectly doable to solve the
equations of motion in the gauge $\ma{A}^-=0$, starting with
eqs.~(\ref{eq:fieldsLC}) as initial conditions.

However, above the forward light-cone, the natural coordinates to
describe a high energy collision is the $(\tau,\x,\eta)$ system. And
consequently, the Fock-Schwinger gauge condition
$\ma{A}^\tau=x^-\ma{A}^++x^+\ma{A}^-=0$ leads to simpler equations of
motion than the light-cone gauge $\ma{A}^-=0$. It is therefore
desirable to apply a gauge transformation to the fields of
eqs.~(\ref{eq:fieldsLC}) in order to satisfy the Fock-Schwinger gauge
condition.

This transformation can be done in two stages. First of all, let us
apply a gauge transformation $\Omega_2\equiv\ma{U}_2$, that changes
the field of the second nucleus before the collision from $A_2^+$ into
a transverse pure gauge $\ma{A}^i=\alpha_2^i$. By doing this, we
arrive at a more symmetric description of the collision, where both
nuclei produce a transverse pure gauge field prior to the collision.
When applied to the fields of eqs.~(\ref{eq:fieldsLC}), this
transformation gives the following fields at $\tau=0^+$~:
\begin{eqnarray}
\partial^-\ma{A}^{+a}(\x)&=&-ig\ma{A}^{i}_{2ab}(\x)\ma{A}_1^{ib}(\x)
\nonumber\\
\ma{A}^{ia}(\x)&=&\ma{A}_1^{ia}(\x)+\ma{A}_2^{ia}(\x)
\nonumber\\
\ma{A}^{\pm a}(\x)&=&0\;.
\label{eq:fieldsLC1}
\end{eqnarray}
The first of eqs.~(\ref{eq:fieldsLC1}) makes an explicit reference to
the components of $\ma{A}^{i}_2$ in the adjoint representation. One
can therefore also rewrite it as a commutator,
\begin{equation}
\partial^-\ma{A}^{+}(\x)=ig[\ma{A}^{i}_{1}(\x),\ma{A}_2^{i}(\x)]\; .
\end{equation}
Note that after this first stage, we are still in the light-cone gauge
$\ma{A}^-=0$, but with a different choice of the residual gauge fixing
compared to eqs.~(\ref{eq:fieldsLC}). Indeed, since $\ma{U}_2$ does
not depend on $x^+$, the gauge transformation generated by $\ma{U}_2$
cannot produce a non-zero $\ma{A}^-$.

As explained in \cite{BlaizM1}, the final step to get the
Fock-Schwinger gauge fields is to perform a gauge transform $\Omega$
such that
\begin{equation}
\ma{A}^{\mu}
=
\Omega \ma{A}_{_{\rm FS}}^{\mu} \Omega^{\dagger}+\frac{i}{g}\Omega \partial^\mu \Omega^{\dagger} \;,
\label{eq:FStrans}
\end{equation}
where the left hand side is the gauge potential of
eqs.~(\ref{eq:fieldsLC1}) in light-cone gauge $\ma{A}^-=0$, and
$\ma{A}_{_{\rm FS}}^\mu$ the gauge potential in Fock-Schwinger gauge.
The $\mu=-$ component of these equations should therefore tell us how
to choose $\Omega$ in order to achieve the desired transformation.
Recalling that $\ma{A}^\pm_{_{\rm FS}}=\pm x^{\pm} \ma{A}^\eta_{_{\rm
    FS}}$, and defining also $\ma{A}^+=x^{+} \ma{A}^\eta$, one then
finds
\begin{equation}
\Omega(\tau,\x)=e^{\frac{ig\tau^2}{2}\ma{A}^\eta(\x)}\; .
\end{equation}
Note that this formula is only valid for very small values of
$\tau>0$, since it has been obtained solely from the knowledge of the
value of the gauge fields at $\tau=0^+$.  Applying then this gauge
transformation to the other components of the gauge potential, we
recover the known results from \cite{KovneMW2,KovneMW1}, that we have
already recalled in eqs.~(\ref{eq:mvfields}).

\section{Small fluctuations at $\tau=0^+$}
\label{sec:fluct}
\subsection{Set up of the problem}
\begin{figure}[htbp]
\begin{center}
\resizebox*{!}{8cm}{\includegraphics{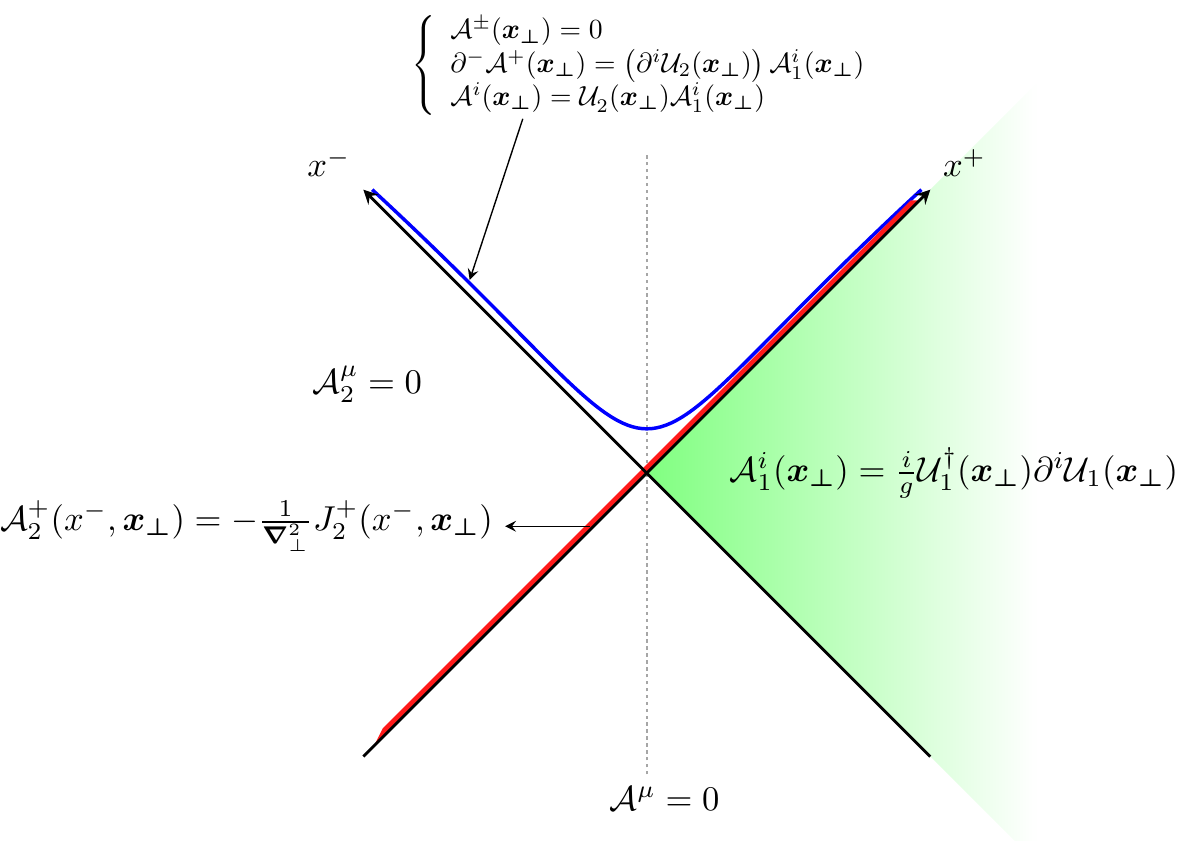}}
\end{center}
\caption{\label{fig:smallfluc}Structure of the background field in the
  light-cone gauge $\ma{A}^-=0$.}
\end{figure}

We now turn to the problem of computing analytically the small
fluctuations $a^\mu$ on top of the background field, with plane wave
initial conditions in the remote past. We will perform most of the
calculation in the same $\ma{A}^-=0$ light-cone gauge that was used in
the previous section for the background field, and the gauge
transformation to obtain finally the fluctuations in the
Fock-Schwinger gauge will be performed at the very end.  The setup of
the problem in this gauge is illustrated in the figure
\ref{fig:smallfluc}, where we indicate the structure of the background
field in each relevant region of space-time.

\begin{figure}[htbp]
\begin{center}
  \resizebox*{12.5cm}{!}{\includegraphics{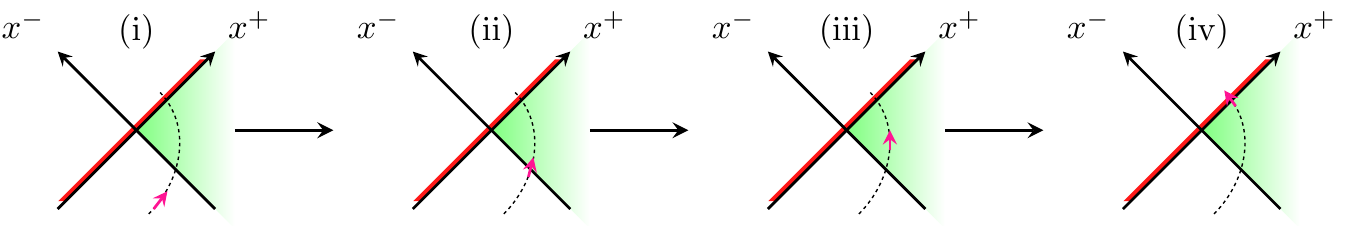}}
\end{center}
\caption{\label{fig:4steps}The four steps in the evolution of a
  fluctuation from $x^0=-\infty$ to the forward light cone. See in
  order the sections \ref{sec:step1}, \ref{sec:step2}, \ref{sec:step3}
  and \ref{sec:step4}.}
\end{figure}
In this calculation, we will consider only the propagation of the
fluctuations on the right part of the space-time diagram in the figure
\ref{fig:smallfluc}, i.e. waves that encounter first the nucleus 1 and
next the nucleus 2. Naturally, there is a second contribution in which
this sequence is reversed, but it is easy to guess it by symmetry at
the end of the calculation. Note that there is no possibility of
cross-talk between these two contributions thanks to causality.
This time evolution of a wave starting at $x^0=-\infty$ can be divided
in four steps, illustrated in the figure \ref{fig:4steps}~:
\begin{itemize}
\item[{\bf i.}] evolution in the region $x^\pm<0$, before the
  fluctuation encounters any of the nuclei,
\item[{\bf ii.}] evolution across the trajectory of the first nucleus,
\item[{\bf iii.}] evolution in the region $x^+>0,x^-<0$, between the two nuclei,
\item[{\bf iv.}] evolution across the trajectory of the second nucleus.
\end{itemize}

The initial plane wave at $x^0=-\infty$ is completely characterized by
a momentum ${\bm k}$, a color $c$, and a polarization $\lambda$, and it
reads
\begin{equation}
a^{\mu{a}}_{{\bm k}\lambda{c}}(x)\equiv \delta_c^a\,\epsilon^\mu_{{\bm k}\lambda}\,e^{ik\cdot x}\; .
\end{equation}
For every momentum ${\bm k}$, there are two physical polarizations, and
we choose their polarization vectors to be mutually orthogonal,
$g_{\mu\nu}\epsilon^{\mu}_{{\bm k}\lambda}\epsilon^{\nu}_{{\bm
    k}\lambda'}=\delta_{\lambda\lambda'}$. In the rest of this
section, we will consistently use the same notation, where the lower
indices are the quantum numbers of the initial plane wave at
$-\infty$, and the upper indices represent its Lorentz and color
structure at the current point $x$.

\subsection{Step {\bf i}: evolution in the backward light cone} 
\label{sec:step1} From now on we will work in the light cone
coordinate system\footnote{We will translate our expressions in the
  $(\tau,\eta,\x)$ coordinate system only in the section
  \ref{sec:tauetacoor}, when the fluctuation reaches the forward
  light-cone.}. The region $x^\pm<0$ located below the trajectories of
the two nuclei is completely trivial, since none of the nuclei has yet
influenced the fluctuation. Thus the equation of motion in this region
are simply the free linearized Yang-Mills equations. In the
$\ma{A}^{-}=0$ gauge, the plane waves in this region read
\begin{align}
a^{i{a}}_{{\bm k} \lambda{c}}(x)=\null&
\delta_c^a\,\epsilon^i_{{\bm k} \lambda}\, e^{ik\cdot x}  &
 a^{+{a}}_{{\bm k} \lambda{c}}(x)=\null&
\delta_{{c}}^{{a}}\,\frac{k^i\epsilon^i_{{\bm k} \lambda}}{k^{-}}\,e^{ik\cdot x}&
a^{-{a}}_{{\bm k} \lambda{c}}(x)=\null&0\; .
\label{eq:freewaves}
\end{align}
Note that the component $\epsilon^+$ of the polarization vector is
constrained by Gauss's law (i.e. the one among the four equations of
motion that does not contain the derivative $\partial^+$, and
therefore acts as a constraint at every value of $x^-$),
\begin{equation}
  \partial_\mu a^{\mu}_{{\bm k} \lambda{c}}=0\; ,
\end{equation}
that requires $k_\mu\epsilon^\mu=0$. The two physical polarizations
are obtained by choosing the transverse polarization vector
$\epsilon^i$, such that $\epsilon^i_{{\bm k} \lambda}\epsilon^i_{{\bm
    k} \lambda'}=\delta_{\lambda\lambda'}$. In the rest of this
section, we will often omit the subscripts ${\bm k}\lambda c$ in the
notation for the fluctuation, in order to lighten a bit the notations.

\subsection{Step {\bf ii}: crossing the trajectory of the first nucleus}
\label{sec:step2}
\begin{figure}[htbp]
\begin{center}
\resizebox*{!}{3cm}{\includegraphics{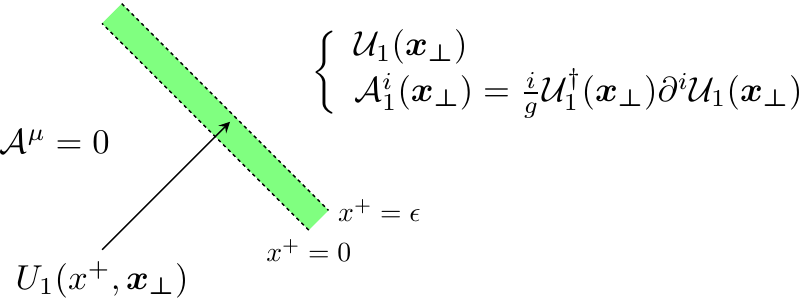}}
\end{center}
\caption{\label{fig:1stcross}Crossing the first nucleus.}
\end{figure}

The first non trivial step of the evolution of the fluctuation is to
cross the trajectory of the first nucleus, on the half-line defined by
$x^+=0, x^-<0$.  Note that here one cannot use the crossing formulas
derived in \cite{GelisM1}, since both the structure of the background
field and the gauge condition for the fluctuation are different.
The first thing to realize is that the fluctuation has a non-zero
$a^+$ component (since we are in the $\ma{A}^-=0$ gauge), that will
induce a precession of the current $J_1^-$ of this nucleus. Therefore,
we need to first consider the current conservation equation for the
nucleus,
\begin{equation}
%\ma{D}^{{ab}}_{1\mu}\ma{F}_1^{\mu\nu{b}}=\null&J_1^{\nu}&
%
\ma{D}^{{ab}}_{1\nu}J_1^{\nu{b}}=0\; .
\label{eq:Jcons}
\end{equation}
For the background field only, the solution reads
\begin{equation}
J_1^{i{a}}=J_1^{+{a}}=0\; ,
\quad
J_1^{-{a}}(x^+,\x)=\ma{U}_{1ab}^{\dagger}(x^+,\x)\rho^{{b}}_1(x^+,\x)\;.
\end{equation} 
To compute the change of this current $j^-$ induced by the component
$a^+$ of the incoming fluctuation, we need to correct
eq.~(\ref{eq:Jcons}) to linear order, which gives~:
\begin{equation}
\partial^+\delta_{{ab}}j_1^{-{b}}=iga^{+}_{{ab}}J_1^{-{b}}\;. 
\end{equation} 
Recalling the fact that $J_1^-$ does not depend on $x^-$, this
equation is solved by
\begin{equation}
j_1^{-{a}}(x)=-ig\,J_{1ab}^{-}(x^+,\x)\,\frac{1}{\partial^+}\,a^{+{b}}(x^+=0,x^-,\x)\;.
\end{equation}
The operator $1/\partial^+$ should be understood as an integration
with respect to $x^-$.  We can now write the linearized Yang-Mills
equations that drive the evolution of the fluctuation across the
infinitesimal region supporting the sources of the first nucleus,
\begin{equation}
\ma{D}^{{ab}}_{\mu}\left(\ma{D}^{\mu{bc}}\,a^{\nu{c}}
-\ma{D}^{\nu{bc}}\,a^{\mu{c}}\right)
-ig\ma{F}^{\nu\mu{ab}}\,a_{\mu}^{{b}}=j^{\nu{a}}
\;.
\label{E:titefluctu}
\end{equation} 
If $0<x^+<\epsilon$ is the range where the sources of this nucleus are
non-zero, then the field strength $\ma{F}^{\nu\mu}$ of the background
gauge potential is identically zero for $x^+>\epsilon$, while inside the
strip $0<x^+<\epsilon$, its only non-zero component is
\begin{equation}
\ma{F}^{-i}=\partial^-\,\ma{A}^{i}_{1}\; .
\end{equation}
This allows the following simplifications of eqs.~(\ref{E:titefluctu})~:
\begin{eqnarray}
&&
-\ma{D}^{{ab}}_{1\mu}(\partial^- a^{\mu{b}})
-ig(\partial^-\ma{A}^{{ab}}_{1\mu})a^{\mu{b}}=j_1^{-{ a}}\nonumber\\
&&
\left(\delta^{{ab}}2\partial^-\partial^+-\ma{D}_1^{i{ac}}\ma{D}_1^{i{cb}}\right)a^{+{b}}
-\partial^{+}\left(\partial^-a^{+{a}}-\ma{D}_1^{i{ab}}a^{i{b}}\right)=0\nonumber\\
&&
\left(2\delta^{{ab}}\partial^-\partial^+-\ma{D}_1^{i{ac}}\ma{D}_1^{i{cb}}\right)a^{j{b}}
-\partial^-\ma{D}_1^{j{ab}}a^{+{b}}
\nonumber\\
&&\qquad\qquad\qquad
+
\ma{D}_1^{i{ac}}\ma{D}_1^{j{cb}}a^{i{b}}
+ig(\partial^-\ma{A}^{j{ab}}_{1})a^{+{b}}=0\;.
\label{eq:fluct1}
\end{eqnarray}
Since we only want to evolve the fluctuation from $x^+=0$ to
$x^+=\epsilon$, we are interested only in the terms of these equations
that can potentially be of order $\epsilon^{-1}$ (i.e. the would
behave as $\delta(x^+)$ in the limit $\epsilon\to 0^+$) and therefore
lead to a finite variation of the fluctuation. Let us recall that
$\ma{A}^i$ has a finite jump in this strip, and therefore the
derivative $\partial^-\ma{A}^i$ behaves as $\epsilon^{-1}$. The
induced current $j^-_1$ behaves similarly, since it is proportional to
the current $J_1^-$.

The first of eqs.~(\ref{eq:fluct1}) has no $\partial^+$ derivative and
can be seen as a constraint at fixed $x^-$: it is nothing but Gauss's
law for the small fluctuation in this gauge. More explicitly, it reads
\begin{align}
\partial^-\left(\partial^- a^{+{a}}-\ma{D}_1^{i{ab}}a^{i{b}}\right)
=\null&2ig(\partial^-\ma{A}^{i{ab}}_{1})a^{i{b}}-j_1^{-{a}}\;,\label{eq:gausslaudepx}
\end{align}
which implies that the combination $\partial^-
a^{+{a}}-\ma{D}^{i{ab}}a^{i{b}}$ changes by a finite amount when going
from $x^+=0$ to $x^+=\epsilon$.  Therefore, the second of
eqs.~(\ref{eq:fluct1}) does not contain any term proportional to
$\epsilon^{-1}$, which implies that $a^{+}$ varies infinitesimally
between $x^+=0$ and $x^+=\epsilon$. We can now simplify the third
equation, by dropping all the terms that are bounded in the limit
$\epsilon\to 0$, which leaves only
\begin{equation}
\partial^-\partial^+a^{j{a}}=-ig\,(\partial^-\ma{A}^{j{ab}}_{1})\,a^{+{b}}\;.
\end{equation}
It is easy to integrate this equation over $x^+$ from $0$ to
$\epsilon$, and since $a^+$ is continuous in the infinitesimal
integration domain, it can be taken out of the integral. This leads to
\begin{equation}
\left[a^{j{a}}\right]_{x^+=\epsilon}
-
\left[a^{j{a}}\right]_{x^+=0}
=
-ig\,\ma{A}^{j{ab}}_{1}(x^+=\epsilon,\x)\,
\frac{1}{\partial^+}a^{+{b}}_0(x)\;,\label{eq:jumponai}
\end{equation}
where we have used again the fact that $\ma{A}^j$ does not depend on
$x^-$.  The subscript $0$ in $a_0^+$ in the right hand side is used to
indicate that this quantity is the free plane wave described in the previous
subsection, in eqs.~(\ref{eq:freewaves}).

In order to be complete, we need to calculate also the
variation of the derivative $\partial^- a^+$. Indeed, although $a^+$
itself varies smoothly, this derivative may have a finite change from
$x^+=0$ to $x^+=\epsilon$. For that, we integrate Gauss's law over
$x^+$ from $0$ to $\epsilon$,
\begin{equation}
\left[\partial^- a^{+{a}}\right]_{x^+=\epsilon}
-
\left[\partial^- a^{+{a}}\right]_{x^+=0}=
\int_0^\epsilon \d x^+\,\left[\ma{D}_1^{i{ab}}\partial^-a^{i{b}}
+ig(\partial^-\ma{A}^{i{ab}}_{1})a^{i{b}}-j_1^{-{a}}\right]\;.
\end{equation}
Using what we have just derived for $a^i$, and using the equation of
motion of the background field, $\ma{D}_{ 1}^{i{ab}}\partial^-\ma{A}_{
  1}^{i{b}}=\null J_1^{-{a}}$, we obtain the following result
\begin{eqnarray}
\left[\partial^- a^{+{a}}\right]_{x^+=\epsilon}
-
\left[\partial^- a^{+{a}}\right]_{x^+=0}
&=&
ig\int_0^\epsilon \d x^+\,(\partial^-\ma{A}^{i{ab}}_{1})\,
\left(a^{i{b}}_0-\frac{\partial^i}{\partial^+}a^{+{b}}_0\right)
\nonumber\\
&=&
ig\,\ma{A}^{i{ab}}_{1}(x^+=\epsilon,\x)
\left(a^{i{b}}_0-\frac{\partial^i}{\partial^+}a^{+{b}}_0\right)\;.
\nonumber\\
&&
\label{eq:jumponap}
\end{eqnarray}
The formulas (\ref{eq:jumponai}) and (\ref{eq:jumponap}), together
with the result that $a^+$ varies smoothly while going from
$x^+=0$ to $x^+=\epsilon$, are the central result of this
subsection. One can also check Gauss's law at this point, which is a
good test of the overall consistency of the solution
\begin{eqnarray}
\left[\partial^-(\partial^-a^{+{a}}-\ma{D}^{i{ab}}a^{i{b}})\right]_{x^+=\epsilon}
\!\!\!&=&\!
ig\left[\ma{D}^{i{ac}}_1\partial^-\ma{A}_1^{i{cb}}\right]_\epsilon
\frac{1}{\partial^+}a_0^{+{b}}
+2ig\left[\partial^-\ma{A}_1^{i{ab}}\right]_\epsilon a_0^{i{b}}
\nonumber\\
&=&
0
\;,\label{eq:gausslaw}
\end{eqnarray}
because the background field $\ma{A}_1^{i}$ is independent of $x^+$ for $x^+\ge 0$.

\subsection{Step {\bf iii}: propagation over the pure gauge $\ma{A}_1^{i}$}
\label{sec:step3}
In this subsection, we consider the evolution of the fluctuation after
it has crossed the trajectory of the first nucleus, and before it
reaches the second one. The results of the previous subsection provide
the initial conditions for this evolution, and the most direct way to
perform the next stage is to write the Green's formula that relates
the value of the fluctuation at any point in the quadrant
$x^+>0,x^-<0$ to this initial data.

Since in this region the Wilson line $\ma{U}(\x)$ depends only on
$\x$, the background field $\ma{A}_1$ is truly a pure gauge, and the
linearized equation of motion (\ref{E:titefluctu}) for the fluctuation
can be written as
 \begin{equation}
\ma{U}^{\dagger}_{1ac}(\x)
\left(g_{\mu\nu}\square-\partial_\mu\partial_\nu\right)
\ma{U}_{1{cb}}(\x)a^{\mu{ b}}(x)
=
0\;,
\end{equation}
which means that the gauge rotated fluctuation
$\widetilde{a}^{\mu{a}}(x)\equiv\ma{U}_{1ab}(\x)\,a^{\mu b}(x)$
propagates over the vacuum. One can easily obtain the following Green's
formula for this free evolution\footnote{The derivation can be found
  in ref. \cite{GelisLV3}.},
\begin{eqnarray}
\widetilde{a}^{\mu}(x)
&=&
i\smash{\int\limits_{y^{+}=0^+}}\d y^-\d^2 \y\; \Big\{
D^{\mu +}_{_R}(x,y)\,\big[\partial_y^{\nu}\widetilde{a}_{\nu}(y)\big]
-\big[\partial^y_{\nu}D^{\mu\nu}_{_R}(x,y)\big]\,\widetilde{a}^{+}(y)
\nonumber\\
&&
\qquad\qquad\qquad\qquad
+D^{\mu i}_{_R}(x,y)\stackrel{\leftrightarrow}{\partial_y^{+}}\widetilde{a}^{i}(y) 
\Big\}
\;,\label{eq:greenf}
\end{eqnarray}
where $D^{\mu\nu}_{_R}$ is the free retarded propagator in the
light-cone gauge $\ma{A}^-=0$, whose expression in momentum space reads~:
\begin{equation}
D^{\mu\nu}_{_R}
=
-\frac{i}{k^2+ik^0\epsilon}\left(g^{\mu\nu}-
\frac{k^{\mu}n^{\nu}+k^{\nu}n^{\mu}}{n.k+i\epsilon}\right)\,,
\end{equation} 
with $n^+=1$, $n^-=n^i=0$. The following formulae will also prove
useful later~:
\begin{eqnarray}
\partial_{\mu}^x\,D^{\mu\nu}_{_R}(x,y)&=&
-i\delta^{\nu+}\theta(x^+-y^+)\,\delta(x^--y^-)\,\delta(\x-\y)\nonumber\\
\partial_{\mu}^y\partial_{\mu}^x\,D^{\mu\nu}_{_R}(x,y)&=&
i\delta(x^+-y^+)\,\delta(x^--y^-)\,\delta(\x-\y)\;.\label{eq:propretprop}
\end{eqnarray}
The Green's formula (\ref{eq:greenf}) is valid everywhere in the
region $x^+>\epsilon,x^-<0$. One can verify that this formula
conserves Gauss's law, as it should. Indeed, since above the
$x^+=\epsilon$ line, $\ma{U}_1$ does not depend on $x^+$, Gauss's law
(\ref{eq:gausslaudepx}) simply becomes
\begin{equation}
\partial^-(\ma{D}_{1ab}^{\mu}\,a^{\mu{b}})=
\ma{U}^{\dagger{ab}}_1(\x)\,\partial^-
\left(\partial^- \widetilde{a}^{+{b}}
-\partial^i\widetilde{a}^{i{b}}\right)=0\;,\label{eq:gausslaw2}
\end{equation}
which implies that $\partial_\mu \widetilde{a}^\mu$ should be
independent of $x^+$. That this is true can easily be checked thanks to
eqs.~(\ref{eq:greenf}) and (\ref{eq:propretprop}).

Some technical results that are necessary in order to calculate
$\widetilde{a}(x)$ for $x^+>\epsilon$ are derived in the appendix
\ref{sec:appendixa}. The results can be written in a more compact form
by introducing modified polarization vectors defined by
\begin{equation}
\widetilde{\epsilon}^i_{{\bm k}\lambda}
\equiv
\left(\delta^{ij}-\frac{2k^ik^j}{k^2}\right)
\epsilon^j_{{\bm k} \lambda}\; .
\end{equation}
These new polarization vectors satisfy
\begin{equation}
k^i\widetilde{\epsilon}^i_{{\bm k}\lambda}
=
-k^i{\epsilon}^i_{{\bm k}\lambda}\; ,\quad
\sum_{i=1,2}\widetilde{\epsilon}^i_{{\bm k}\lambda}\widetilde{\epsilon}^i_{{\bm k}\lambda'}
=\sum_{i=1,2}{\epsilon}^i_{{\bm k}\lambda}{\epsilon}^i_{{\bm k}\lambda'}=\delta_{\lambda\lambda'}\; .
\end{equation}

Thanks to eq.~(\ref{eq:greenf}), we obtain~:
\begin{eqnarray}
\widetilde{a}^{i{a}}_{{\bm k} \lambda{c}}(x)
&=& 
e^{ik^+x^-}\intp\; e^{i\p\cdot\x}\;
\Big[e^{i\frac{p^2}{2k^+}x^+}\Big(\delta^{ij} -\frac{2p^ip^j}{p^2}\Big)
\nonumber\\
&&\qquad\qquad
+2p^i\Big(\frac{p^j}{p^2}+\frac{k^j}{k^2}\Big)\Big]\;
\ma{\widetilde{U}}^{{ac}}_1(\p+\ka)\;\widetilde{\epsilon}^j_{{\bm k} \lambda}
\nonumber\\
\widetilde{a}^{+{a}}_{{\bm k} \lambda{c}}(x)&=&
2k^+e^{ik^+x^-}\intp\; e^{i\p\cdot\x}\;
\Big[e^{i\frac{p^2}{2k^+}x^+}\frac{p^i}{p^2}\nonumber\\
&&\qquad\qquad
-\Big(\frac{p^i}{p^2}+\frac{k^i}{k^2}\Big)\Big]\;
\ma{\widetilde{U}}^{{ac}}_1(\p+\ka)\;\widetilde{\epsilon}^i_{{\bm k} \lambda}\;,
\end{eqnarray}
where we denote $k\equiv|\ka|$, $p\equiv|\p|$, and where
$\ma{\widetilde{U}}_1(\ka)$ is the Fourier transform of
$\ma{U}_1(\x)$,
\begin{equation}
\ma{\widetilde{U}}_1(\ka)
=
\int \d^2\x\;e^{-i\ka\cdot\x}\;\ma{U}_1(\x)\; .
\end{equation}
Undoing the gauge rotation $\ma{U}_1$ to go back to $a^\mu$
gives~:
\begin{eqnarray}
a^{i{a}}_{{\bm k} \lambda{c}}(x)
&=& 
e^{ik^+x^-}\ma{U}^{{ab}{\dagger}}_1(\x) \intp \;e^{i\p\cdot\x}\;
\Big[e^{i\frac{p^2}{2k^+}x^+}\Big(\delta^{ij} -\frac{2p^ip^j}{p^2}\Big)
\nonumber\\
&&\qquad\qquad
+2p^i\Big(\frac{p^j}{p^2}+\frac{k^j}{k^2}\Big)\Big]\;
\ma{\widetilde{U}}^{{bc}}_1(\p+\ka)\;\widetilde{\epsilon}^j_{{\bm k} \lambda}\;,
\label{eq:aistep3}\\
a^{+{a}}_{{\bm k} \lambda{c}}(x)
&=&
2k^+e^{ik^+x^-}\ma{U}^{{ab}{\dagger}}_1(\x)\intp \;e^{i\p\cdot\x}\;
\Big[e^{i\frac{p^2}{2k^+}x^+}\frac{p^i}{p^2}
\nonumber\\
&&\qquad\qquad
-\Big(\frac{p^i}{p^2}+\frac{k^i}{k^2}\Big)\Big]\;
\ma{\widetilde{U}}^{{bc}}_1(\p+\ka)\;\widetilde{\epsilon}^i_{{\bm k} \lambda}\;.
\label{eq:apstep3}
\end{eqnarray}
These formulas are valid in the entire quadrant $x^+>\epsilon, x^-<0$.
The last step of the evolution is now to let the fluctuation cross the
trajectory of the second nucleus. In the appendix \ref{app:checks}, we
perform several consistency checks on the formulas
(\ref{eq:aistep3}-\ref{eq:apstep3}).

\subsection{Step {\bf iv}: crossing the trajectory of the second nucleus}
\label{sec:step4}
\begin{figure}[htbp]
\begin{center}
\resizebox*{!}{3cm}{\includegraphics{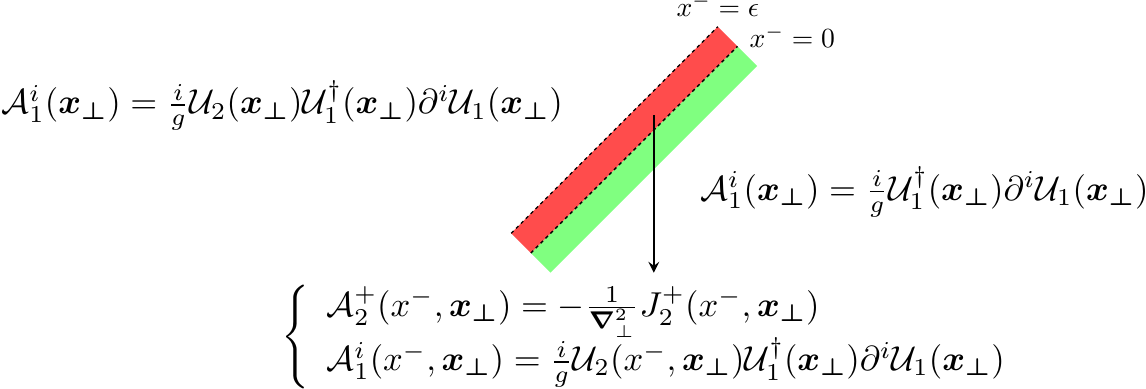}}
\end{center}
\caption{\label{fig:2ndcross}Crossing the second nucleus.}
\end{figure}
The propagation of the fluctuation through the second nucleus is very
similar to the situation studied in \cite{GelisM1}. The only
difference is that in \cite{GelisM1}, the gauge potential of the
nucleus had only an $\ma{A}^+$ component, proportional to
$\delta(x^-)$. In the present situation, the nucleus also has a
non-zero $\ma{A}^i$. Despite this important difference, the
calculation can be done in a very similar way as in \cite{GelisM1}. In
particular, the fact that we are in the $\ma{A}^-=0$ light-cone gauge
prevents any precession of the color current $J_2^+$ of the nucleus.
The linearized Yang-Mills equations for the fluctuation therefore take
a simpler form, without any source term in the right hand
side\footnote{${\cal D}^\mu\equiv \partial^\mu-ig{\cal A}^\mu$ is the
  generic notation for the covariant derivative in light-cone gauge,
  while ${\cal D}_1^i\equiv\partial^i-ig{\cal A}_1^i$ and ${\cal
    D}_2^+\equiv\partial^+-ig{\cal A}_2^+$ are built specifically with
  the fields ${\cal A}_1^i$ and ${\cal A}_2^+$ respectively.},
\begin{equation}
\ma{D}^{{ab}}_{\mu}\left(\ma{D}^{\mu}_{bc}\,a^{\nu{c}}-\ma{D}^{\nu}_{bc}\,a^{\mu{c}}
\right)-ig\ma{F}^{\nu\mu}_{ab}\,a_{\mu}^{{b}}=0
\;.
\label{E:titefluctustep4}
\end{equation} 
For the component $\nu=-$, using the fact that $\mathcal{A}_1^i$ does
not depend on $x^+$, this gives Gauss's law~:
\begin{equation}
\partial^-\left(\partial^-a^{+{a}}-\ma{D}_{1ab}^{i}\,a^{i{b}}\right)=0\;.
\label{eq:gauss3}
\end{equation}
For $\nu=i$, keeping only terms that have a singular behavior in
$\epsilon^{-1}$, we get 
\begin{equation}
\partial^-\ma{D}_{2ab}^{+}\,a^{i{b}}=0\; ,
\end{equation} 
which can be solved by
\begin{equation}
a^{i{a}}_{{\bm k} \lambda{c}}(x^-=\epsilon,x^+,\x)
=
\ma{U}_{2ab}(x^-,\x)\;
a^{i{b}}_{{\bm k} \lambda{c}}(x^-=0,x^+,\x)\;.
\label{eq:cross2-1}
\end{equation}
Finally, for $\nu=+$, the equation of motion reads
\begin{eqnarray}
&&
\left(2\partial^-\ma{D}_{2ab}^{+}-\ma{D}_{1ac}^{i}\ma{D}_{1cd}^{i}\right)\,a^{+{b}}
-\left(\partial^-\ma{D}_{2ab}^{+}\, a^{+{b}}-\ma{D}_{1ac}^{i}\ma{D}_{2cb}^{+}\,a^{i{b}}
\right)\nonumber\\
&&\qquad\qquad\qquad
+gf^{{abc}}(\partial^+\ma{A}_{1c}^{i}-ig\ma{A}_{2cd}^{+}\ma{A}_{1d}^{i}
-\partial^i\ma{A}_{2c}^{+})\,a^{i{b}}=0\;.
\end{eqnarray}
Since $\partial^+\ma{A}_1^{i{c}}=ig\ma{A}_{2cd}^{+}\ma{A}_1^{i{d}}$ in
the strip $0<x^-<\epsilon$, the previous equation can be simplified
into
\begin{equation}
\ma{D}_{2ab}^{+}\,a^{+{b}}
=
ig(\partial^i\ma{A}_{2ab}^{+})\,\frac{1}{\partial^-}\,a^{i{b}}\;.
\end{equation}
The solution of this equation is known (see \cite{GelisM1}), and
agrees with Gauss's law,
\begin{eqnarray}
&&a^{+{a}}_{{\bm k} \lambda{c}}(x^-=\epsilon,x^+,\x)
=
\ma{U}_{2ab}(\x)\,a^{+{b}}_{{\bm k} \lambda{ c}}(x^-=0,x^+,\x)\nonumber\\
&&\qquad\qquad\qquad\qquad
+(\partial^i\ma{U}_{2ab}(\x))\,\frac{1}{\partial^-}a^{i{b}}_{{\bm k} \lambda{c}}(x^-=0^-,x^+,\x)\; .
\label{eq:cross2-2}
\end{eqnarray}
It turns out in the end that the equations (\ref{eq:cross2-1}) and
(\ref{eq:cross2-2}) are identical to the crossing formulas of
\cite{GelisM1}, despite the presence of a non-vanishing
$\ma{A}_1^i$. Eqs.~(\ref{eq:cross2-1}) and (\ref{eq:cross2-2}) provide
the value of the fluctuation in the light-cone gauge $\ma{A}^-=0$ on
the right branch of the light-cone, just after the collision.  Our
next task will be to convert these expressions into the Fock-Schwinger
gauge.

\subsection{From light-cone gauge to Fock-Schwinger gauge}
Like for the background field itself, the first stage in this process
is to go back to the situation where both nuclei are described by
transverse pure gauges before the collision. To do so, we first
perform a gauge transform $\ma{U}_2^\dagger$ which trivially affects
the small fluctuations~: 
\begin{equation}
a^\mu\quad\rightarrow\quad \alpha^\mu\equiv\ma{U}_2\; a^\mu\; \ma{U}_2^{\dagger}\;.
\end{equation} 
After this first gauge transformation, the fluctuations on the right
branch of the light-cone therefore read
\begin{eqnarray}
\alpha^{+{a}}_{{\bm k} \lambda{c}}(x^-=\epsilon)
&=&
a^{+{a}}_{{\bm k} \lambda{c}}(x^-=0)
-ig\ma{A}_{2ab}^{i}(\x)\frac{1}{\partial^-}a^{i{b}}_{{\bm k} \lambda{c}}(x^-=0)
\nonumber\\
\alpha^{i{a}}_{{\bm k} \lambda{c}}(x^-=\epsilon)
&=&
a^{i{a}}_{{\bm k} \lambda{c}}(x^-=0)
\;.
\label{eq:step4}
\end{eqnarray}
In order to check Gauss's law at this point, one should recall that
after this transformation the transverse background gauge potential is
now $\mathcal{A}^{i}=\mathcal{A}_1^{i}+\mathcal{A}_2^{i}$, and
therefore the covariant derivative in eq.~(\ref{eq:gauss3}) should
be modified accordingly.

To go to the Fock-Schwinger gauge, we must perform one last gauge
transformation $W$, in analogy with the transformation of
eq.~(\ref{eq:FStrans}) for the background field. The crucial point to
note is that $W$ must differ from $\Omega$ (where $\Omega$ is the
gauge transformation used to transform the background field into the
Fock-Schwinger gauge), since the fluctuation depends on $\eta$. But
since the fluctuation is small compared to the background field, $W$
should be close to $\Omega$, $W\equiv\Omega+ig\omega$, with $\omega$ of
order unity.  The action of this gauge transformation on the
background field and on the small fluctuations can be split as
follows~:
\begin{eqnarray}
\ma{A}^{\mu}
&=&
\Omega \ma{A}_{_{\rm FS}}^{\mu} \Omega^{\dagger}+\frac{i}{g}\Omega \partial^\mu \Omega^{\dagger}
\nonumber\\
\alpha^\mu
&=&
\Omega\alpha^\mu_{_{\rm FS}}\Omega^\dagger
+
\Omega\partial^\mu\omega
+
ig\Big(\omega\ma{A}^\mu_{_{\rm FS}}\Omega^\dagger
-\Omega\ma{A}^\mu_{_{\rm FS}}\omega\Big)\; ,
\end{eqnarray}
where the light-cone gauge quantities are in the left hand side, and
the Fock-Schwinger gauge quantities carry a FS subscript. When
$\tau\to 0^+$, the gauge rotation $\Omega$ goes to $1$, and therefore
the transformation of the fluctuation simply becomes
\begin{equation}
\alpha^{\mu a}
=
\alpha^{\mu a}_{_{\rm FS}}
+
\mathcal{D}_{_{\rm FS}}^{\mu ab}\,\omega^b\; ,
\label{eq:FStrans-a}
\end{equation}
where $\mathcal{D}_{_{\rm FS}}$ is the covariant derivative
constructed with the background field near $\tau=0^+$ in the
Fock-Schwinger gauge. Like in the case of the background field, we
will obtain $\omega$ by requesting that $\alpha^-=0$.

Like for the background field, let us parameterize the $\pm$
components of the fluctuations as follows,
\begin{equation}
\alpha_{_{\rm FS}}^{\pm}\equiv\pm x^{\pm}\;\alpha^{\eta}_{_{\rm FS}}\quad,\qquad
\alpha^{+}\equiv x^+\;\alpha^{\eta}\; .
\end{equation} 
From the component $\mu=-$ of eq.~(\ref{eq:FStrans-a}), we get 
\begin{equation}
x^-\,\alpha^{{a}\eta}_{_{\rm FS}}
=\partial^-\omega^{{a}}
-ig\mathcal{A}_{_{\rm FS}}^{-{ab}}\omega^{{b}}\; ,
\end{equation} 
and in terms of the coordinates $\tau,\eta$, we obtain
\begin{equation}
\tau\,\alpha^{{a}\eta}_{_{\rm FS}}=
\partial_{\tau}\omega^{{a}}+\frac{1}{\tau}\partial_\eta \omega^{{a}}+ig\tau\mathcal{A}_{_{\rm FS}}^{{ab}\eta}\omega^{{b}} \;.
\end{equation}
Injecting this into the $\mu=+$ equation and using $x^+=\tau
e^\eta/\sqrt{2}$ gives
\begin{equation}
\omega^{{a}}(\tau,\eta,\x)
=
\frac{1}{\sqrt{2}}\int_{0}^{\tau}\d\tau'\;e^{-\eta}\;\alpha^{+ a}(\tau',\eta,\x)\;,
\label{eq:omega}
\end{equation}
where we recall that $\alpha^+$ is given by eqs.~(\ref{eq:step4}) and
(\ref{eq:apstep3}).  In terms of this $\omega$, the gauge
transformation formulas for the fluctuation can also be written as
\begin{equation}
\alpha_{_{\rm FS}}^{i{a}}
=
 \alpha^{i{a}}-\ma{D}^{i}_{ab}\,\omega^{{b}}
\quad,\quad
\alpha_{_{\rm FS}}^{\eta{a}}
=
\frac{1}{2}\alpha^{\eta{a}}
+
\frac{ig}{2}\,\ma{A}^{\eta}_{ab}\,\omega^{{b}}
+
\frac{1}{\tau^2}\partial_\eta\, \omega^{{a}} 
\label{eq:flucfsint}\;.
\end{equation}
(We recall that all the fields and fluctuations without the FS
subscript are in the light-cone gauge $\ma{A}^-=0$.) It turns out that
the terms $\ma{D}^{i}_{{ab}}\omega^{{b}}$ and
$\frac{ig}{2}\ma{A}^{\eta}_{ab}\,\omega^{{b}}$ will not contribute at
lowest non-zero order in $\tau$.

\subsection{Expression in terms of the conjugate momentum to $\eta$} 
\label{sec:tauetacoor}
Eqs.~(\ref{eq:aistep3}-\ref{eq:apstep3}), (\ref{eq:step4}), and
(\ref{eq:omega}-\ref{eq:flucfsint}) provide the value in
Fock-Schwinger gauge, just above the forward light-cone, for any
fluctuation that started as a free plane wave in the remote past.
However, for practical uses of these fluctuations in heavy ion
collisions, the numerical implementation will be performed on a
lattice that has a fixed extent in the rapidity $\eta$, which is more
appropriate for the description of a system in rapid expansion in the
longitudinal direction.

Anticipating the use of this system of coordinates, it would be
desirable to have an ensemble of fluctuations labeled by a quantum
number which is the conjugate momentum of rapidity (that we shall
denote $\nu$ in the following), instead of the conjugate momentum
$k_z$ of the Cartesian longitudinal coordinate $z$. Obviously, since
the two sets of fluctuations both are a basis of the vector space of
all fluctuations, there must be a linear transformation to obtain one
from the other. The transformation that goes from the basis of
fluctuations labeled by $k_z$ to the basis of fluctuations labeled
by $\nu$ reads
\begin{equation}
\alpha_{\ka \nu \lambda c}^\mu(x)= \int_{-\infty}^{+\infty} dy\; e^{i\nu y}\;
\alpha_{{\bm k}\lambda c}^\mu(x)\; ,
\label{eq:y-to-nu}
\end{equation}
where $y\equiv \log(k^+/k^-)/2$ is the momentum rapidity. Indeed,
using the fact that the problem is invariant under boosts in the
longitudinal direction, the fluctuation $\alpha_{{\bm k}\lambda
  c}^\mu(x)$ must depend on $y$ and $\eta$ only via the difference
$y-\eta$. Changing the integration variable $y$ in favor of $y'\equiv
y-\eta$, we readily see that the $\eta$ dependence of the left hand
side of eq.~(\ref{eq:y-to-nu}) is of the form $\exp(i\nu\eta)$. This
shows that this transformation indeed leads to fluctuations that have
a well defined conjugate momentum $\nu$ to the rapidity $\eta$.

We have now all the ingredients to compute the final form of the
fluctuations\footnote{The following integral is also useful in order to
  perform the integration over the rapidity $y$,
  \begin{equation*}
    \int \d y\,e^{i\nu y}\,e^{ y-\eta} 
    e^{i\frac{\tau p^2}{2k}e^{\eta-y}}
    =-i e^{i\nu \eta}\left(\frac{ p}{k}\right)^{i\nu+1}\Gamma(-1-i\nu)e^{\frac{\nu\pi}{2}}  \left(\frac{\tau p}{2}\right)^{1+i\nu}\;.
  \end{equation*}}.
After a straightforward but tedious calculation, we obtain
the following formulas for the fluctuations and the corresponding electrical fields
\begin{eqnarray}
\alpha^{_{\rm R}\,i a}_{_{\rm FS}\,\ka \nu\lambda c}(\tau,\eta,\x)
&=&
F_{\ka\nu\lambda c}^{+,ia}(\tau,\eta,\x)
\nonumber\\
e^{_{\rm R}\,i a}_{_{\rm FS}\,\ka \nu\lambda c}(\tau,\eta,\x)
&=&
-i\nu \;F_{\ka\nu\lambda c}^{+,ia}(\tau,\eta,\x)\nonumber\\
\alpha_{_{\rm FS}\,\ka \nu\lambda c}^{_{\rm R}\, \eta a}(\tau,\eta,\x)
&=&
\frac{1}{2+i\nu}\,\ma{D}^{iab}_{_{\rm FS}}\; F_{\ka\nu\lambda c}^{+,ib}(\tau,\eta,\x)
\nonumber\\
e^{_{\rm R}\,\eta a}_{_{\rm FS}\,\ka \nu\lambda c}(\tau,\eta,\x)
&=&
-\ma{D}^{iab}_{_{\rm FS}}\;F_{\ka\nu\lambda c}^{+,ib}(\tau,\eta,\x)
\; ,
\label{eq:centralresult}
\end{eqnarray}
(the superscript R indicates that we have only the contribution due to
the wave that propagates on the right of the light-cone) where we
denote
\begin{eqnarray}
&&
F_{\ka\nu\lambda c}^{+,ia}(\tau,\eta,\x)
\equiv
\Gamma(-i\nu)\,e^{+\frac{\nu\pi}{2}}
e^{i\nu\eta}\,\ma{U}^{\dagger}_{1ab}(\x)\,\widetilde{\epsilon}^j_{{\bm k} \lambda}
\nonumber\\
&&\quad\times
\intp\; e^{i\p\cdot\x}\;
\ma{\widetilde{U}}_{1bc}(\p+\ka)
\left(\frac{p_\perp^2\tau}{2k_\perp}\right)^{+i\nu}
\Big[\delta^{ji} -\frac{2p^j_\perp p^i_\perp}{p^2_\perp}\Big]
\; .
\label{eq:F1}
\end{eqnarray}
In order to write the $\eta$ components of the gauge potential and of
the electrical field as a covariant derivative acting on the function
$F_{\ka\nu\lambda c}^{+,ia}(\tau,\eta,\x)$, we have used the following
identity,
\begin{equation}
\ma{D}_{1ab}^i\,\ma{U}^\dagger_{1bc}=\ma{D}_{2ab}^i\,\ma{U}^\dagger_{2bc}=0\; .
\label{eq:DUdag}
\end{equation}

These formulas (to be completed by eqs.~(\ref{eq:centralresult2}),
that give the result when the fluctuations have propagated on the
other side of the light-cone) are the central result of this
work. They provide analytical expressions for fluctuations with plane
wave initial conditions in the remote past, after they have propagated
over the classical background field created in a heavy ion collision,
in the Fock-Schwinger gauge and with a set of quantum numbers
appropriate for a discretization on a lattice with a fixed spacing in
the rapidity $\eta$. Note that in the first formula, we have written
$\alpha^\eta$ (with the Lorentz index up). The corresponding
$\alpha_\eta$ (with the Lorentz index down) is obtained by multiplying
by $-\tau^2$ and therefore becomes very small when $\tau\to 0^+$.

In the limit $\tau\to 0^+$, the fluctuations of the potentials
and electrical fields behave in the same manner as their counterpart
in the background field, except for the transverse electrical field
$e^i$. The background field has a transverse electrical field
$\ma{E}^i$ that vanishes like $\tau^2$, while its fluctuations $e^i$
go to a non-zero limit when $\tau\to 0^+$ for all the modes
$\nu\not= 0$.

The dependence of these fluctuations on the classical background field
is known explicitly, and is entirely contained in the Wilson lines
that appear in the function $F_{\ka\nu\lambda
  c}^{+,ia}(\tau,\eta,\x)$, and in the covariant derivatives that
appear in some of the eqs.~(\ref{eq:centralresult}). From this
function, it is easy to obtain all the components of the fluctuations
and the corresponding electrical fields thanks to
eqs.~(\ref{eq:centralresult}). The numerical evaluation of
$F_{\ka\nu\lambda c}^{+,ia}(\tau,\eta,\x)$ is rather straightforward,
since it only involves a pair of Fourier transforms\footnote{Even if
  the formulas proposed in ref.~\cite{DusliGV1} were not affected by
  the caveat raised in the introduction, they would be more difficult
  to evaluate numerically since they require that one solves a large
  eigenvalue problem. Moreover, the separation between the physical
  and unphysical modes was problematic in ref.~\cite{DusliGV1}. In
  contrast, the approach followed in the present paper gives directly
  the physical modes.}.

\subsection{Contribution from the propagation on the left}
Eqs.~(\ref{eq:centralresult}) have been derived by evolving the small
fluctuations in the right part of the light cone (crossing first the
nucleus $1$, and then crossing the nucleus $2$). To this contribution
should be added the contribution obtained by the other ordering of the
encounters with the two nuclei, i.e. when the fluctuation propagates
on the left side of the light-cone. This extra contribution is
completely independent from the one we have just calculated, since by
causality they cannot talk to each other.

This new contribution can be obtained by repeating the same steps as
the ones employed so far, but now working in the $\ma{A}^+=0$ gauge.
This leads to~:
\begin{eqnarray}
\alpha^{_{\rm L}\,i a}_{_{\rm FS}\,\ka \nu\lambda c}(\tau,\eta,\x)
&=&
F_{\ka\nu\lambda c}^{-,ia}(\tau,\eta,\x)
\nonumber\\
e^{_{\rm L}\,i a}_{_{\rm FS}\,\ka \nu\lambda c}(\tau,\eta,\x)
&=&
i\nu \;F_{\ka\nu\lambda c}^{-,ia}(\tau,\eta,\x)\nonumber\\
\alpha_{_{\rm FS}\,\ka \nu\lambda c}^{_{\rm L}\, \eta a}(\tau,\eta,\x)
&=&
-\frac{1}{2-i\nu}\,\ma{D}^{iab}_{_{\rm FS}}\; F_{\ka\nu\lambda c}^{-,ib}(\tau,\eta,\x)
\nonumber\\
e^{_{\rm L}\,\eta a}_{_{\rm FS}\,\ka \nu\lambda c}(\tau,\eta,\x)
&=&
\ma{D}^{iab}_{_{\rm FS}}\;F_{\ka\nu\lambda c}^{-,ib}(\tau,\eta,\x)
\; ,
\label{eq:centralresult2}
\end{eqnarray}
where the superscript L indicates that this is the partial wave that
has propagated on the left part of the light-cone, and where we now
denote
\begin{eqnarray}
&&
F_{\ka\nu\lambda c}^{-,ia}(\tau,\eta,\x)
\equiv
\Gamma(+i\nu)\,e^{-\frac{\nu\pi}{2}}
e^{i\nu\eta}\,\ma{U}^{\dagger}_{2ab}(\x)\,
\widetilde{\epsilon}^j_{{\bm k} \lambda}
\nonumber\\
&&\quad\times
\intp\; e^{i\p\cdot\x}\;
\ma{\widetilde{U}}_{2bc}(\p+\ka)
\left(\frac{p_\perp^2\tau}{2k_\perp}\right)^{-i\nu}
\Big[\delta^{ji} -\frac{2p^j_\perp p^i_\perp}{p^2_\perp}\Big]
\; .
\label{eq:F2}
\end{eqnarray}

\subsection{Complete result}
Let us finally add up the results of eqs.~(\ref{eq:centralresult}) and
(\ref{eq:centralresult2}), in order to obtain the complete value of
the fluctuations just above the forward light cone~:
\begin{eqnarray}
\alpha^{i a}_{_{\rm FS}\,\ka \nu\lambda c}(\tau,\eta,\x)
&=&
F_{\ka\nu\lambda c}^{+,ia}(\tau,\eta,\x)+F_{\ka\nu\lambda c}^{-,ia}(\tau,\eta,\x)
\nonumber\\
e^{i a}_{_{\rm FS}\,\ka \nu\lambda c}(\tau,\eta,\x)
&=&
-i\nu \;\Big(F_{\ka\nu\lambda c}^{+,ia}(\tau,\eta,\x)-F_{\ka\nu\lambda c}^{-,ia}(\tau,\eta,\x)\Big)
\; .\nonumber\\
\alpha_{_{\rm FS}\,\ka \nu\lambda c}^{\eta a}(\tau,\eta,\x)
&=&
\ma{D}^{iab}_{_{\rm FS}}\; 
\Big(
\frac{F_{\ka\nu\lambda c}^{+,ib}(\tau,\eta,\x)}{2+i\nu}
-
\frac{F_{\ka\nu\lambda c}^{-,ib}(\tau,\eta,\x)}{2-i\nu}
\Big)
\nonumber\\
e^{\eta a}_{_{\rm FS}\,\ka \nu\lambda c}(\tau,\eta,\x)
&=&
-\ma{D}^{iab}_{_{\rm FS}}\;
\Big(F_{\ka\nu\lambda c}^{+,ia}(\tau,\eta,\x)-F_{\ka\nu\lambda c}^{-,ia}(\tau,\eta,\x)\Big)\; ,
\nonumber\\
&&
\label{eq:finalresult}
\end{eqnarray}
with the functions $F_{\ka\nu\lambda c}^{+,ia}$ and $F_{\ka\nu\lambda
  c}^{-,ia}$ defined in eqs.~(\ref{eq:F1}) and (\ref{eq:F2})
respectively. These formulas are the analogue for gluons of the
eq.~(14) of Ref.~\cite{GelisKL1}, that had been derived for the quark
mode functions. Analogous formulas are also known for leptons in the
QED electromagnetic background created by two colliding electrical
charges~\cite{BaltzM1}.

\subsection{Various checks}
The most obvious check one can perform is that the fluctuations given
by eqs.~(\ref{eq:finalresult}) satisfy the equations of motion at
lowest order in $\tau$, i.e. in an infinitesimal domain $\tau\ll
Q_s^{-1}$ above the forward light-cone,
\begin{align}
\frac{1}{\tau}\partial_{\tau}
\left(\frac{1}{\tau}\partial_{\tau}\right)\,\alpha_{_{\rm FS}\,\eta}^{{ a}}
+\frac{i\nu}{\tau^2}\mathcal{D}_{_{\rm FS}}^{i{ a}{ b}}\,\alpha_{_{\rm FS}}^{i{ b}}
=\null&0\;,&
\left[\frac{1}{\tau}\partial_{\tau}
\left(\tau\partial_{\tau}\right)
+\frac{\nu^2}{\tau^2}\right]\,\alpha_{_{\rm FS}}^{i{ a}}=\null&0\;.
\end{align}

A more stringent test is to check whether Gauss's law is still
satisfied, because it involves a delicate interplay between the
background field and the fluctuation. We indeed find that
\begin{equation}
\partial_\eta e^\eta_{_{\rm FS}}-\ma{D}_{_{\rm FS}}^ie^i_{_{\rm FS}}=0\;.
\end{equation}
Note that the terms in $ig\ma{E}_{_{\rm FS}}^\eta \alpha_{_{\rm
    FS}\,\eta}-ig\ma{E}_{_{\rm FS}}^i \alpha_{_{\rm FS}}^i$, that are
normally part of the Gauss's law for a fluctuation in the
Fock-Schwinger gauge, are of higher order in $\tau$, and do not play a
role here.  Moreover, the terms in $\tau^{\pm i \nu}$ in the solution
satisfy independently Gauss's law, as they should since they have
evolved independently on each side of the light-cone.

Finally, one can again compute the inner product of two of the
fluctuations we have obtained in (\ref{eq:finalresult}). The term in
$\alpha_{_{\rm FS}\,\eta}^*e^\eta_{_{\rm FS}}-e^{\eta*}_{_{\rm
    FS}}\alpha_{_{\rm FS}\,\eta}$ does not contribute at lowest order
in $\tau$, and therefore the inner product simply reads
\begin{eqnarray}
&&
\big(
\alpha^{_{\rm FS}}_{{\bm k}_\perp \nu \lambda{c}}
\big|
\alpha^{_{\rm FS}}_{{\bm k}_\perp' \nu' \lambda'{d}}
\big)
=
i\!\!\int\!\! \d^2 \x \d \eta\, 
\Big(
\alpha^{i{{a}*}}_{_{\rm FS}\,{\bm k}_\perp \nu \lambda{c}}(\tau,\eta,\x)
e^{i{a}}_{_{\rm FS}\,{\bm k}_\perp' \nu' \lambda'{d}}(\tau,\eta,\x)
\nonumber\\
&&
\qquad\qquad\qquad\qquad
-
e^{i{a}*}_{_{\rm FS}\,{\bm k}_\perp \nu \lambda{c}}(\tau,\eta,\x)
\alpha^{i{a}}_{_{\rm FS}\,{\bm k}_\perp' \nu' \lambda'{d}}(\tau,\eta,\x)\Big)\;.
\end{eqnarray}
Using the fact that $\nu|\Gamma(i\nu)|^2(e^{\pi\nu}-e^{-\pi\nu})=2\pi$, we find
\begin{align}
\big(
\alpha^{_{\rm FS}}_{{\bm k}_\perp \nu \lambda{c}}
\big|
\alpha^{_{\rm FS}}_{{\bm k}_\perp' \nu' \lambda'{d}}
\big)=
4\pi(2\pi)^3\delta(\nu-\nu')\delta(\ka-\ka')\delta_{\lambda\lambda'}\delta_{{ cd}}\;.
\label{eq:pstaueta}
\end{align}
To prove that this is indeed the correct answer, let us recall what
this inner product should be before we made the transformation $k_z\to
\nu$,
\begin{align}
\big(
\alpha^{_{\rm FS}}_{{\bm k} \lambda{c}}
\big|
\alpha^{_{\rm FS}}_{{\bm k}' \lambda'{d}}
\big)=
2|k^0|(2\pi)^3\delta({\bm k}-{\bm k}')\delta_{\lambda\lambda'}\delta_{{ cd}}\;.
\end{align}
Using
\begin{align}
2|k^0|(2\pi)^3\delta({\bm k}-{\bm k}')
=
2(2\pi)^3\delta(\ka-\ka')\delta(y-y')\;,
\end{align}
and applying the transformation $\int \d y dy'\;e^{i(\nu y-\nu'y')}$
to the right hand side, we find that the inner product in the basis
where we use the quantum number $\nu$ instead of $k_z$ should indeed
be given by eq.~(\ref{eq:pstaueta}).

\subsection{Numerical implementation}
\label{sec:nums}
From eqs.~(\ref{eq:finalresult}), (\ref{eq:F1}) and (\ref{eq:F2}) it
is obvious that the most difficult and time consuming part in
evaluating numerically these fields is the computation of the
functions $F_{\ka\nu\lambda c}^{+,ia}$ and $F_{\ka\nu\lambda
  c}^{-,ia}$. Let us list here the main steps in their computation~:
\begin{itemize}
\item[{\bf i.}] Compute the Wilson lines ${\cal U}_{1,2}$ that
  represent the color charge content of the two colliding nuclei in
  the McLerran-Venugopalan model. This is very easy for the SU(2)
  gauge group, and a little more involved for SU(3).
\item[{\bf ii.}] Compute the Fourier transform (over the transverse
  coordinate $\x$) $\widetilde{\cal U}_{1,2}$ of these Wilson
  lines. Since space is discretized on a lattice, this is a discrete
  Fourier transform, for which there are some very efficient
  algorithms\footnote{Naive algorithms for the discrete Fourier
    transform of an array of size $L$ scale as $L^2$, while the
    efficient implementations scale as $L\log(L)$.}.
\item[{\bf iii.}] The integration over $\p$ in
  eqs.~(\ref{eq:F1}) and (\ref{eq:F2}) is also a discrete Fourier
  transform.
\end{itemize}
This is the work that needs to be done to compute one of the mode
functions, with given quantum numbers $\ka,\nu,\lambda,c$. When
computing a generic perturbation to the gauge potential, one must sum
over all these mode functions with random weights. Since the
$\eta,\nu$ dependence of the mode functions is in $\exp(i\nu\eta)$,
the sum over the index $\nu$ can also be viewed as a discrete Fourier
transform. For computing $N_{\rm conf}$ configurations of these
fluctuating fields, on a lattice that has $L\times L$ sites in the
transverse direction, and $N$ sites in the $\eta$ direction, the
computational cost scales as
\begin{equation}
N_{\rm conf}\times N\log(N)\times L^4\log(L)\; .
\label{eq:cost1}
\end{equation}
This is the estimate for a straightforward implementation. A more careful
examination of how the various steps of the calculation depend on each
other leads to a better algorithm, whose cost scales as
\begin{equation}
N\log(N)\times L^4\times(A\log(L)+BN_{\rm conf})\; ,
\label{eq:cost2}
\end{equation}
with $A$ and $B$ two constants. For large $L$ and/or $N_{\rm conf}$, this
is significantly  better than (\ref{eq:cost1}).

The only subtlety arises when discretizing the first order differential
operators $\ma{D}^{i},\partial^{i}$ and the corresponding momenta such
as $p^{i},k^{i}$ that enter in (\ref{eq:finalresult}). They can be
discretized either as backward or forward finite differences. The
choice between the two is arbitrary, and is completely determined by
what kind of discretization is chosen for the derivatives in the
linearized Gauss law.

\section{Conclusions and outlook}
\label{sec:concl}
In this work, we have performed an explicit calculation of the small
fluctuations that must be superimposed to the classical CGC field in
order to resum the unstable modes of the Yang-Mills equations. The
calculation has been done from first principles, by solving the
evolution equation for small fluctuations on top of the classical
background field, with plane wave initial conditions in the remote
past.

Although the intermediate steps of the evolution are done in the
light-cone gauge $\ma{A}^-=0$, the final results are given in the
Fock-Schwinger gauge $\ma{A}^\tau=0$. Moreover, they are also given in
terms of the quantum number $\nu$, Fourier conjugate of the rapidity
$\eta$, which is conserved when the background field in independent of
rapidity. Fluctuations expressed in terms of $\nu$ are also more
suitable for a numerical implementation on a lattice with a fixed
spacing in $\eta$. Our final formulas, eqs.~(\ref{eq:finalresult}),
are valid just after the collision, at proper times $\tau\ll
Q_s^{-1}$. By construction, they obey the linearized Yang-Mills
equations with the correct initial condition at $x^0=-\infty$, and
satisfy Gauss's law. They have also been checked to satisfy the
expected orthonormality conditions, when we compute the inner product
defined in eq.~(\ref{eq:centralps}).

These formulas also turn out to be very compact, and are
straightforward to implement numerically. They will be essential in
the study of the behavior at early times of the strong color fields
produced in high energy heavy ion collisions.  Indeed, it has been
noticed a long time ago that certain modes are subject to the Weibel
instability and therefore have an exponential growth in time, and that
this effect may play a crucial role in the isotropization and
thermalization (the basic idea being that it could be fast thanks to
the exponential growth of these modes). In the CGC framework, these
modes first appear at NLO, where they can give corrections that become
sizable (as large as the LO contribution) in a short time. It is
therefore important to compute these NLO corrections in order to
reliably describe the behavior at early times of the fields produced
in a collision. And as explained in the introduction, one needs the
mode functions derived in this paper in order to perform this
calculation.

Note that with mode functions that differ from the ones derived here,
one would still trigger the Weibel instabilities since any randomly
chosen fluctuations are likely to have a non-zero projection on some
of the unstable modes. However, the timescale for the growth of the
fields depends on the initial amplitude of the fluctuations, in
particular their amplitude relative to that of the background field
(this is were the dependence on the coupling constant $g$ comes from,
since the background is of order $1/g$ while the fluctuations are of
order $1$). This timescale also depends on the relative amplitude of
the various mode functions, since the growth rate of the Weibel
instability depends on the quantum numbers $\ka$ and $\nu$. Therefore,
in order to assess correctly the early time evolution of the system,
it is necessary to use the mode functions calculated here, that have
been constructed in order to guarantee an accurate result up to NLO.

\section*{Acknowledgements}
We thank J. Berges, J.-P. Blaizot, K. Dusling, S. Jeon, L. McLerran,
S. Schlichting, and R. Venugopalan for useful discussions related to
this work. FG thanks the Nuclear Theory group at BNL, as well as the
University of Cape Town and NITheP, for their hospitality and
financial support at various stages of this work. This research is
supported by the Agence Nationale de la Recherche project
11-BS04-015-01.

\appendix

\section{Useful formulas in the derivation of
  (\ref{eq:aistep3}-\ref{eq:apstep3})} 
\label{sec:appendixa}
This appendix provides some formulas that are useful in order to
compute the fluctuation $\widetilde{a}^\mu$ via the Green's formula
(\ref{eq:greenf})~:
\begin{eqnarray}
&&
i\int\limits_{y^{+}=0^+}\d y^-\d^2 \y\;
D^{j i}_{_R}(x,y)\stackrel{\leftrightarrow}{\partial_y^{+}}\,e^{iky}\,\alpha(\y)
\nonumber\\
&&\qquad\qquad
=
\delta^{ij}e^{ik^+x^-}\intp\; \widetilde{\alpha}(\p+\ka)\,e^{i\p\cdot\x}\,e^{i\frac{p^2}{2k^+}x^+}\;,
\end{eqnarray}
\begin{eqnarray}&&
i\int\limits_{y^{+}=0^+}\d y^-\d^2 \y\; D^{i +}_{_R}(x,y)\,e^{iky}\,\alpha(\y)
\nonumber\\
&&\quad
=
-ie^{ik^+x^-}\intp\; \widetilde{\alpha}(\p+\ka)\,e^{i\p\cdot\x}\,\frac{p^i}{p^2}\,
\Big(1-e^{i\frac{p^2}{2k^+}x^+}\Big)\;,
\end{eqnarray}
\begin{eqnarray}&&
i\int\limits_{y^{+}=0^+}\d y^-\d^2 \y\;
D^{+ i}_{_R}(x,y)\stackrel{\leftrightarrow}{\partial_y^{+}}\,e^{iky}\,\alpha(\y)
\nonumber\\
&&\quad
=
2k^+\,e^{ik^+x^-}\intp\; \widetilde{\alpha}(\p+\ka)\,e^{i\p\cdot\x}\,\frac{p^i}{p^2}\,
\Big(1-e^{i\frac{p^2}{2k^+}x^+}\Big)\;,
\end{eqnarray}
\begin{eqnarray}&&
i\int\limits_{y^{+}=0^+}\d y^-\d^2 \y\; 
\Big[
\partial^y_{\mu}D^{+\mu}_{_R}(x,y)
\Big]\,
\widetilde{a}^{+}(y)
=
-\widetilde{a}^{+}(x^+=0^+)\; ,
\end{eqnarray}
\begin{eqnarray}&&
i\int\limits_{y^{+}=0^+}\d y^-\d^2 \y\;
 D^{+ +}_{_R}(x,y)\,e^{iky}\,\alpha(\y)
\nonumber\\
&&\quad
=
i\,e^{ik^+x^-}\intp\;\widetilde{\alpha}(\p+\ka)\,e^{i\p\cdot\x}\,\frac{2k^+}{p^2}\,
\Big(1-e^{i\frac{p^2}{2k^+}x^+}\Big)\;,
\end{eqnarray}
where $\widetilde{\alpha}$ denotes the transverse Fourier transform of
the function $\alpha(\y)$,
\begin{equation}
\widetilde{\alpha}(\ka)\equiv \int \d^2\y\; e^{-i\ka\cdot\y}\;\alpha(\y)\; .
\end{equation}
In order to obtain these formulas, one should replace the free
retarded propagator by its Fourier representation, and perform the
integral over the energy in the complex plane thanks to the theorem of
residues.

\section{Various checks of eqs.~(\ref{eq:aistep3}-\ref{eq:apstep3})}
\label{app:checks}
First of all, one can check that the Gauss's law, given in
eq.~(\ref{eq:gausslaw2}), is indeed satisfied by
eqs.~(\ref{eq:aistep3}-\ref{eq:apstep3}). From these formulas, one can
readily see that (\ref{eq:aistep3}-\ref{eq:apstep3})
\begin{align}
\partial_\mu\widetilde{a}^{\mu{a}}_{{\bm k} \lambda{c}}(x)=\null&
2 e^{ik^+x^-}e^{-i\ka.\x}\left(i\frac{\partial^i\partial^ik^j\epsilon^j_{{\bm k} \lambda}}{k^2}
-\partial^i\epsilon^i_{{\bm k} \lambda}\right)\ma{U}^{{ac}}_1(\x)\;.
\end{align}
Using eq.~(\ref{eq:DUdag}), which also implies
\begin{equation}
\ma{D}^{i}_{1ac}(\x)\,\ma{A}^{i}_{1cb}(\x)
=
\frac{i}{g}\ma{U}^{\dagger}_{1ac}(\x)\,\partial^i\partial^i\,\ma{U}_{1cb}(\x)
\end{equation}
and
$\ma{D}^{\mu}_{1ac}(\x)\,a^{\mu{c}}=\partial_\mu\widetilde{a}^{\mu{a}}$,
we obtain (\ref{eq:gausslaw2}).

Another non-trivial check is to compute the inner product on the
$x^-=0$ surface (i.e. before the fluctuation traverses the trajectory
of the second nucleus). This inner product reads
\begin{equation}
\big(a_{{\bm k} \lambda { c}}\big|a_{{\bm k}' \lambda' { d}}\big)
=
-i\int\limits_{x^-=0}\d^2{\bm x}_{\perp}\d x^+\;
a_{{\bm k} \lambda { c}}^{i{ a}*}\,
\stackrel{\leftrightarrow}{\partial^-}\,
a_{{\bm k}' \lambda' { d}}^{i{ a}}\;.
\end{equation}
That is calculated by dividing the integration on $x^+$ in three
pieces: $-\infty<x^+<0$, $0<x^+<\epsilon$, and
$\epsilon<x^+<+\infty$. Note that the second range gives a finite
contribution, despite its infinitesimal size, because
$\partial^-\ma{A}^i_1$ behaves as $\epsilon^{-1}$. Doing this
calculation is tedious but straightforward. Using the following
identities,
\begin{eqnarray}
&&
k^+\delta(k^+-k^{'+})=k^-\delta(k^--k^{'-})=|k^0|\delta(k_z-k'_{z})
\qquad\mbox{(for $k^2=k'^2=0$)}
\nonumber\\
&&
\intp\;
 \ma{\widetilde{U}}^{\dagger}_{1cb}(\p+\ka)\ma{\widetilde{U}}_{1bd}(\p+\ka')
=
\delta_{{cd}}\,(2\pi)^2\delta(\ka-\ka')\; ,
\end{eqnarray}
a (somewhat lengthy) calculation gives
\begin{eqnarray}
\big(a_{{\bm k} \lambda { c}}\big|a_{{\bm k}' \lambda' { d}}\big)
&=&
\delta_{\lambda\lambda'}\,\delta_{{ cd}}\,(2\pi)^32|k^0|\,\delta({\bm k}-{\bm k}')
\nonumber\\
&&
+4g\,\frac{\epsilon^-_{{\bm k} \lambda}\epsilon^-_{{\bm k}' \lambda'}}{kk'}
\int\limits_0^{\epsilon}\d x^+ \d^2 {\bm x}_\perp\;  e^{i(\ka-{\bm k}'_\perp)\cdot{\bm x}_\perp}\;
J_{1cd}^{-}(x^+,{\bm x}_\perp)\;.
\nonumber\\
&&
\label{eq:psxpxm}
\end{eqnarray}
The right hand side of this inner product has a somewhat unexpected
term, proportional to the integral of the color current of the first
nucleus. As we shall see now, this term is correct and is the
consequence of the fact that we are in a gauge where the incoming wave
induces a change in this current (because it has a non-zero $a^+$
component that induces a precession of $J_1^-$). This induced current
enters in the equation of motion for the fluctuation itself, and
produces this extra term in the inner product.

Quite generally, in a gauge where such an induced current may appear,
the equation of motion of the fluctuation reads
\begin{equation}
\ma{D}^{{ab}}_{\mu}\left(\ma{D}^{\mu{bc}}\,a^{\nu{c}}_{\k\lambda c}
-\ma{D}^{\nu{bc}}_{\beta}\,a^{\mu{c}}_{\k\lambda c}\right)
-ig\ma{F}^{\nu}_{\mu ab}\,a^{\mu{b}}_{\k\lambda c}=j^{\nu{a}}_{\k\lambda c}
\;.
\label{E:titefluctu1}
\end{equation} 
Because of the induced current in the right hand side, the variation
of the inner product between two (locally space-like) surfaces
$\Sigma_1$ and $\Sigma_2$ may be non zero. More specifically, one has
\begin{equation}
(a_{{\bm k} \lambda { c}}|a_{{\bm k}' \lambda' { d}})_{_{\Sigma_2}}
-
(a_{{\bm k} \lambda { c}}|a_{{\bm k}' \lambda' { d}})_{_{\Sigma_1}}
=
\int_\Omega d^4x\;
\Big(
a_{{\bm k} \lambda { c}}^{+ a *}j^{-{a}}_{\k'\lambda' d}
-
j^{+{a}*}_{\k\lambda c}a_{{\bm k}' \lambda' { d}}^{+ a}
\Big)\; ,
\label{eq:consJ}
\end{equation}
where $\Omega$ is the 4-dimensional domain comprised between the
surfaces $\Sigma_1$ and $\Sigma_2$. In other words, the inner product
is conserved only if there are no induced currents between the two
surfaces on which it is calculated. 

In the situation of interest to us here, the surface $\Sigma_1$ is
entirely located below the backward light-cone, and the surface
$\Sigma_2$ is the plane $x^-=0$ (just below the trajectory of the
second nucleus). The first term in the right hand side of
eq.~(\ref{eq:psxpxm}) is nothing but $(a_{{\bm k} \lambda {
    c}}|a_{{\bm k}' \lambda' { d}})_{_{\Sigma_1}}$. In the right hand
side of eq.~(\ref{eq:consJ}), one can perform analytically the
integral over $x^-$, which gives the extra term in the right hand side
of eq.~(\ref{eq:psxpxm}).

\section{Vacuum solutions} 
\label{sec:appendixb}
In this appendix, we derive the transformation from the $\ma{A}^-=0$
gauge to the Fock-Schwinger gauge in the case of fluctuations
propagating in the vacuum (i.e. when the background field is zero).
In this situation, the fluctuations in light-cone gauge are completely
trivial, of the form
\begin{align}
\alpha^i=\epsilon^i\;e^{ik\cdot x}\quad
,\quad
\alpha^+=\frac{k^i\epsilon^i}{k^-}\;e^{ik\cdot x}\;.
\end{align}
The transformation to Fock-Schwinger gauge can be done via
eq.~(\ref{eq:flucfsint}), simplified here thanks to the absence of
background field
\begin{align}
\alpha_{_{\rm FS}}^{i}=\alpha^{i}-\partial^i\omega
\quad,\quad
\alpha_{_{\rm FS}}^{\eta}=\frac{1}{2}\alpha^{\eta}
+\frac{1}{\tau^2}\partial_\eta \omega\label{eq:flucfsvac}\;,
\end{align}
with
\begin{align}
\alpha^{\eta}=\frac{\alpha^+}{x^+}
\quad,\quad
\omega=\int_0^{\tau} \d \tau'\; \frac{\tau'}{2}\;\alpha^{\eta}\;.
\end{align}
One obtains easily the following explicit expression for $\omega$,
\begin{align}
\omega(\tau,\eta,{\bm x}_\perp)
=
-i\frac{k^i\epsilon^i}{k_\perp^2}\;
e^{-i\ka\cdot\x}\;\frac{e^{y-\eta}\,\big(e^{ik_\perp\tau\cosh(y-\eta)}-1\big)}{\cosh(y-\eta)}\; .
\end{align}
This leads to
\begin{eqnarray}
\alpha^i_{_{\rm FS}}(\tau,\eta,\x)
&=&
\epsilon^j\,e^{-i\ka\cdot\x}\;
\Big[
\delta^{ij}\,e^{ik_\perp\tau\cosh(y-\eta)}
\nonumber\\
&&\qquad
-
\frac{k^ik^j}{k_\perp^2}\frac{e^{y-\eta}\,\big(e^{ik_\perp\tau\cosh(y-\eta)}-1\big)}{\cosh(y-\eta)}
\Big]\; ,
\label{eq:ai-vac}
\\
\alpha_{_{\rm FS}}^{\eta}(\tau,\eta,\x)
&=&
\frac{k^i\epsilon^i}{k_\perp\tau}
\,e^{-i\ka\cdot\x}\;\frac{1}{\cosh(y-\eta)}
\Big[
e^{ik_\perp\tau\cosh(y-\eta)}
\nonumber\\
&&\qquad\qquad\qquad\qquad
+i\frac{e^{ik_\perp\tau\cosh(y-\eta)}-1}{k_\perp\tau\cosh(y-\eta)}
\Big]\; .
\label{eq:aeta-vac}
\end{eqnarray}
The final step is to go from the quantum number $k_z$ to the Fourier
conjugate of rapidity, $\nu$. This is achieved by a Fourier transform
of the $y$ dependence,
\begin{equation}
f(y)\quad\to\quad g(\nu)\equiv \int dy\;e^{iy\nu}\;f(y)\; .
\end{equation}
After this transformation, eqs.~(\ref{eq:ai-vac}) and
(\ref{eq:aeta-vac}) become respectively
\begin{eqnarray}
\alpha^i_{_{\rm FS}}(\tau,\eta,\x)
&=&
\pi e^{-\frac{\pi\nu}{2}}\epsilon^j\,e^{i(\nu\eta-\ka\cdot\x)}\;
\Big[
i\Big(\delta^{ij}-\frac{k^ik^j}{k_\perp^2}\Big)H_{i\nu}^{(1)}(k_\perp\tau)
\nonumber\\
&&\qquad\qquad\qquad\qquad
-\nu\frac{k^ik^j}{k_\perp^2}\int_0^\tau \frac{d\tau'}{\tau'}\,H_{i\nu}^{(1)}(k_\perp\tau')
\Big]
\label{eq:ai-vac-nu}
\end{eqnarray}
and
\begin{eqnarray}
\alpha_{_{\rm FS}\,\eta}(\tau,\eta,\x)
&=&
\pi e^{-\frac{\pi\nu}{2}}k^j\epsilon^j\,e^{i(\nu\eta-\ka\cdot\x)}
\int_0^\tau {d\tau'}\,{\tau'}\,H_{i\nu}^{(1)}(k_\perp\tau')\; ,
\label{eq:aeta-vac-nu}
\end{eqnarray}
where $H_{i\nu}^{(1)}$ is the Hankel function defined in terms of the
Bessel functions as $H_{i\nu}^{(1)}\equiv J_{i\nu}+i Y_{i\nu}$. One
can check that the two vacuum solutions\footnote{The issue in
  ref.~\cite{DusliGV1}, that we raised in the introduction, is only
  with the solutions in the presence of a non-trivial background
  field, which is the case of interest in heavy ion collisions. Note
  that in ref.~\cite{BergeBSV1}, the authors took as initial
  conditions the {\sl vacuum} fluctuations of ref.~\cite{DusliGV1} and
  rescaled them in order to obtain a prescribed gluon occupation
  number at their starting time (which is much larger than
  $Q_s^{-1}$).} found in ref.~\cite{DusliGV1} (eqs.~(79) and (84)) can
be rewritten as linear combinations of the present
eqs.~(\ref{eq:ai-vac-nu}-\ref{eq:aeta-vac-nu}).

Note that if we had started from the vacuum plane wave solutions in the
$\ma{A}^+=0$ light-cone gauge instead, we would have
$\alpha^{'\eta}=-\alpha^{'-}/x^-$ and the function $\omega'$ (we
denote with a prime all the quantities obtained from this alternate
starting point) that defines the transformation to the Fock-Schwinger
gauge would be
\begin{align}
\omega'(\tau,\eta,{\bm x}_\perp)
=
-i\frac{k^i\epsilon^{'i}}{k_\perp^2}\;
e^{-i\ka\cdot\x}\;\frac{e^{\eta-y}\,\big(e^{ik_\perp\tau\cosh(y-\eta)}-1\big)}{\cosh(y-\eta)}\; .
\end{align}
Consequently, the vacuum fluctuations in the Fock-Schwinger gauge
would be replaced by
\begin{eqnarray}
\alpha^{'i}_{_{\rm FS}}(\tau,\eta,\x)
&=&
\epsilon^{'j}\,e^{-i\ka\cdot\x}\;
\Big[
\delta^{ij}\,e^{ik_\perp\tau\cosh(y-\eta)}
\nonumber\\
&&\qquad
-
\frac{k^ik^j}{k_\perp^2}\frac{e^{\eta-y}\,\big(e^{ik_\perp\tau\cosh(y-\eta)}-1\big)}{\cosh(y-\eta)}
\Big]\; ,
\label{eq:ai-vac-+}
\\
\alpha_{_{\rm FS}}^{'\eta}(\tau,\eta,\x)
&=&
-\frac{k^i\epsilon^{'i}}{k_\perp\tau}
\,e^{-i\ka\cdot\x}\;\frac{1}{\cosh(y-\eta)}
\Big[
e^{ik_\perp\tau\cosh(y-\eta)}
\nonumber\\
&&\qquad\qquad\qquad\qquad
+i\frac{e^{ik_\perp\tau\cosh(y-\eta)}-1}{k_\perp\tau\cosh(y-\eta)}
\Big]\; .
\label{eq:aeta-vac-+}
\end{eqnarray}
Recalling that the polarization vectors $\epsilon^i$ in the
$\ma{A}^-=0$ gauge and $\epsilon^{'i}$ in the $\ma{A}^+=0$ gauge are
related by
\begin{equation}
\epsilon^{'i}=\left(\delta^{ij}-2\frac{k^ik^j}{k_\perp^2}\right)\,\epsilon^j\; ,
\end{equation}
it is trivial to see that $\alpha^\mu_{_{\rm FS}}$ and
$\alpha^{'\mu}_{_{\rm FS}}$ differ by a gauge transformation that
does not depend on $\tau$. In other words, they correspond to two different
ways of fixing the residual gauge freedom in the Fock-Schwinger gauge.
This is why the vacuum limit ($\ma{U}_{1,2}\to 1$) of
eqs.~(\ref{eq:finalresult}) does not give precisely
eqs.~(\ref{eq:ai-vac-nu}-\ref{eq:aeta-vac-nu}). Indeed,
eqs.~(\ref{eq:finalresult}) correspond to fixing this residual gauge
freedom independently on the two branches of the light-cone.

%\bibliography{biblio}

\begin{thebibliography}{10}

\bibitem{Teane1}
{D. Teaney}, Prog. Part. Nucl. Phys. {\bf 62}, 451 (2009).

\bibitem{Teane2}
{D. Teaney}, arXiv:0905.2433.

\bibitem{Ollit2}
{J.Y. Ollitrault}, J. Phys. Conf. Ser. {\bf 312}, 012002 (2011).

\bibitem{OllitG1}
{J.Y. Ollitrault, F. Gardim}, Nucl. Phys. {\bf A} {\bf 904--905}, 75c (2013).

\bibitem{PolicSS1}
{G. Policastro, D.T. Son, A.O. Starinets}, Phys. Rev. Lett. {\bf 87}, 081601
  (2001).

\bibitem{BlaizG1}
{J.P. Blaizot, F. Gelis}, Nucl. Phys. A{\bf 750} 148 (2005).

\bibitem{Lappi6}
{T. Lappi}, Int. J. Mod. Phys. {\bf E20}, 1 (2011).

\bibitem{GelisIJV1}
{F. Gelis, E. Iancu, J. Jalilian-Marian, R. Venugopalan}, Ann. Rev. Part. Nucl.
  Sci. {\bf 60}, 463 (2010).

\bibitem{GriboLR1}
{L.V. Gribov, E.M. Levin, M.G. Ryskin}, Phys. Rept. {\bf 100}, 1 (1983).

\bibitem{MuellQ1}
{A.H. Mueller, J-W. Qiu}, Nucl. Phys. {\bf B} {\bf 268}, 427 (1986).

\bibitem{IancuLM3}
{E. Iancu, A. Leonidov, L.D. McLerran}, Lectures given at Cargese Summer School
  on QCD Perspectives on Hot and Dense Matter, Cargese, France, 6-18 Aug 2001,
  hep-ph/0202270.

\bibitem{IancuV1}
{E. Iancu, R. Venugopalan}, Quark Gluon Plasma 3, Eds. R.C. Hwa and X.N. Wang,
  World Scientific, hep-ph/0303204.

\bibitem{Gelis15}
{F. Gelis}, arXiv:1211.3327 [hep-ph].

\bibitem{KovneMW2}
{A. Kovner, L.D. McLerran, H. Weigert}, Phys. Rev. {\bf D} {\bf 52}, 6231
  (1995).

\bibitem{LappiM1}
{T. Lappi, L.D. McLerran}, Nucl. Phys. {\bf A} {\bf 772}, 200 (2006).

\bibitem{KrasnV1}
{A. Krasnitz, R. Venugopalan}, Phys. Rev. Lett. {\bf 84}, 4309 (2000).

\bibitem{KrasnV3}
{A. Krasnitz, R. Venugopalan}, Nucl. Phys. {\bf B} {\bf 557}, 237 (1999).

\bibitem{KrasnNV2}
{A. Krasnitz, Y. Nara, R. Venugopalan}, Phys. Rev. Lett. {\bf 87}, 192302
  (2001).

\bibitem{Lappi1}
{T. Lappi}, Phys. Rev. {\bf C} {\bf 67}, 054903 (2003).

\bibitem{FukusG1}
{K. Fukushima, F. Gelis}, Nucl. Phys. {\bf A} {\bf 874}, 108 (2012).

\bibitem{RomatV1}
{P. Romatschke, R. Venugopalan}, Phys. Rev. Lett. {\bf 96}, 062302 (2006).

\bibitem{RomatV2}
{P. Romatschke, R. Venugopalan}, Eur. Phys. J. {\bf A} {\bf 29}, 71 (2006).

\bibitem{RomatV3}
{P. Romatschke, R. Venugopalan}, Phys. Rev. {\bf D} {\bf 74}, 045011 (2006).

\bibitem{KunihMOST1}
{T. Kunihiro, B. Muller, A. Ohnishi, A. Schafer, T.T. Takahashi, {\bf A}
  Yamamoto}, Phys. Rev. {\bf D} {\bf 82}, 114015 (2010).

\bibitem{Mrowc2}
{S. Mrowczynski}, Phys. Lett. {\bf {\bf B} 214}, 587 (1988).

\bibitem{Mrowc3}
{S. Mrowczynski}, Phys. Lett. {\bf {\bf B} 314}, 118 (1993).

\bibitem{RomatS1}
{P. Romatschke, M. Strickland}, Phys. Rev. {\bf D} {\bf 68}, 036004 (2003).

\bibitem{RomatS2}
{P. Romatschke, M. Strickland}, Phys. Rev. {\bf D} {\bf 70}, 116006 (2004).

\bibitem{RebhaS1}
{A.K. Rebhan, D. Steineder}, Phys. Rev. {\bf D} {\bf 81}, 085044 (2010).

\bibitem{RebhaSA1}
{A.K. Rebhan, M. Strickland, M. Attems}, Phys. Rev. {\bf D} {\bf 78}, 045023
  (2008).

\bibitem{ArnolLM1}
{P. Arnold, J. Lenaghan, G.D. Moore}, JHEP {\bf 0308}, 002 (2003).

\bibitem{ArnolLMY1}
{P. Arnold, J. Lenaghan, G.D. Moore, L.G. Yaffe}, Phys. Rev. Lett. {\bf 94},
  072302 (2005).

\bibitem{ArnolM3}
{P. Arnold, G.D. Moore}, Phys. Rev. {\bf D} {\bf 76}, 045009 (2007).

\bibitem{KurkeM1}
{A. Kurkela, G.D. Moore}, JHEP {\bf 1112}, 044 (2011).

\bibitem{KurkeM2}
{A. Kurkela, G.D. Moore}, JHEP {\bf 1111}, 120 (2011).

\bibitem{BodekR1}
{D. Bodeker, K. Rummukainen}, JHEP {\bf 0707}, 022 (2007).

\bibitem{AttemRS1}
{M. Attems, A. Rebhan, M. Strickland}, Phys. Rev. {\bf D 87}, 025010 (2013).

\bibitem{GelisV2}
{F. Gelis, R. Venugopalan}, Nucl. Phys. {\bf A} {\bf 776}, 135 (2006).

\bibitem{GelisLV3}
{F. Gelis, T. Lappi, R. Venugopalan}, Phys. Rev. {\bf D} {\bf 78}, 054019
  (2008).

\bibitem{GelisLV2}
{F. Gelis, T. Lappi, R. Venugopalan}, Int. J. Mod. Phys. E {\bf 16}, 2595
  (2007).

\bibitem{DusliGV1}
{K. Dusling, F. Gelis, R. Venugopalan}, Nucl. Phys. {\bf A} {\bf 872}, 161
  (2011).

\bibitem{FukusGM1}
{K. Fukushima, F. Gelis, L. McLerran}, Nucl. Phys. {\bf A} {\bf 786}, 107
  (2007).

\bibitem{DusliEGV1}
{K. Dusling, T. Epelbaum, F. Gelis, R. Venugopalan}, Nucl. Phys. {\bf A} {\bf
  850}, 69 (2011).

\bibitem{EpelbG1}
{T. Epelbaum, F. Gelis}, Nucl. Phys. {\bf A} {\bf 872}, 210 (2011).

\bibitem{DusliEGV2}
{K. Dusling, T. Epelbaum, F. Gelis, R. Venugopalan}, Phys. Rev. {\bf D} {\bf
  86}, 085040 (2012).

\bibitem{BlaizM1}
{J.P. Blaizot, Y. Mehtar-Tani}, Nucl. Phys. {\bf A} {\bf 818}, 97 (2009).

\bibitem{KovneMW1}
{A. Kovner, L.D. McLerran, H. Weigert}, Phys. Rev. {\bf D} {\bf 52}, 3809
  (1995).

\bibitem{GelisM1}
{F. Gelis, Y. Mehtar-Tani}, Phys. Rev. {\bf D} {\bf 73}, 034019 (2006).

\bibitem{GelisKL1}
{F. Gelis, K. Kajantie, T. Lappi}, Phys. Rev. C. {\bf 71}, 024904 (2005).

\bibitem{BaltzM1}
{A.J. Baltz, L.D. McLerran}, Phys. Rev. {\bf C} {\bf 58}, 1679 (1998).

\bibitem{BergeBSV1}
{J. Berges, K. Boguslavski, S. Schlichting, R. Venugopalan}, arXiv:1303.5650.

\end{thebibliography}
%\bibliographystyle{unsrt}

\end{document}